\documentclass[11pt,fullpage]{article}
\usepackage{graphicx,latexsym}
\usepackage{longtable}

\newcommand{\greeksym}[1]{{\usefont{U}{psy}{m}{n}#1}}

\newcommand{\uomega}{\mbox{\greeksym{w}}}
\newcommand{\uxi}{\mbox{\greeksym{x}}}
\newcommand{\uphi}{\mbox{\greeksym{f}}}
\newcommand{\uPfi}{\mbox{\greeksym{F}}}

\newcommand{\beq}{\begin{equation}}
\newcommand{\eeq}{\end{equation}}
\newcommand{\beqn}{\begin{eqnarray}}
\newcommand{\eeqn}{\end{eqnarray}}

\begin{document}

\title{Quasiresonance}
  
\author{Antonia Ruiz$^{1,2}$ and Eric J. Heller$^{1,3}$\\
\small{$^{1}$Department of Physics, Harvard University, Cambridge, MA 02138}\\
\small{$^{2}$Departamento de F\'\i sica Fundamental y Experimental, Electr\'onica y Sistemas,}\\
\small{Universidad de La Laguna, La Laguna 38203, Tenerife, Spain}\\
\small{$^{3}$Department of Chemistry and Chemical Biology, Harvard University, Cambridge, MA 02138}
}
\maketitle

\begin{abstract}

The concept of quasiresonance was  introduced in connection with inelastic collisions 
between one atom and a vibro-rotationally excited diatomic molecule. In its original form, the 
collisions induce  {\sl quasiresonant}  transfer of energy between the 
internal degrees of freedom of the diatom: there is a surprisingly accurate low order rational value for the 
ratio of the changes in the vibrational and rotational   classical actions, provided the vibrational and rotational frequencies of the diatom are approximately related by low order rational values, and the collision was longer that the rotational period of  the molecule. In this paper  we show that quasiresonance  can be extended to many other processes and systems, and that it may be understood in terms of the adiabatic invariance theory and the method of averaging.  

\end{abstract}

\section{Introduction}

The term quasiresonant vibration-rotation energy transfer (QRVR) was first introduced
\cite{exp1,clastraj} in the context of low velocity inelastic collision processes between one atom and 
a vibro-rotationally excited diatomic molecule.
These are transitions which dominate all other transitions over a significant range 
of initial conditions and occur when there is a highly specific and efficient inelastic transfer
of energy between the internal vibrational and rotational degrees of freedom of the diatom.

The first experimental evidence of QRVR \cite{exp1} was reported in a level-resolved study of 
vibrotationally inelastic collisions between $Li_2^*$ and noble gas atoms. The experimental results show 
remarkably sharp, strong peaks in the vibro-rotational inelastic rate constant at very specific high 
initial rotational states with values that substantially exceed the purely rotationally inelastic 
rate constant. A extremely good correlation between the change in the rotational $(j)$ and vibrational 
$(v)$ quantum numbers of the diatomic at the most important peak of the final rotational states was 
also observed, with a very strong propensity rule $\Delta j = - 4 \Delta v$.

Since then QRVR transfer has been observed in a wide variety of collision species and interaction 
potentials at different energy ranges \cite{exp2,exp3}.
It has been suggested as a possible relaxation mechanism of rotationally excited light molecules,
such as $H_2$ molecules formed in laboratory traps or interstellar clouds, and good candidate to provide
a rotational inversion mechanism in rotationally pumped lasers at ultracold temperatures \cite{exp1}.
Also, a similar quasiresonance energy transfer mechanism has been shown  to control the
process of predissociation for the most weakly bound state of van der Waals complexes \cite{prediss}.

In general, the qualitative features that characterize the quasiresonant vibration-rotation energy transfer are a 
large vibrationally inelastic cross section with very narrow final rotational state distributions,  a strong correlation
\beq\label{quas1}
\Delta j = -\left ( \frac{p}{q} \right )\Delta v,
\eeq
with $p/q$ a rational fraction involving small integers.  The correlation becomes stronger at higher initial rotational states 
and lower collision velocities.
Finally, experimental results  obtained for different collision species also indicate a remarkable
insensitivity to both the initial vibrational state and the nature of the interaction potential \cite{exp1,exp2,exp3}.

These striking properties of the quasiresonant vibro-rotational energy transfer effect have stimulated a
number of theoretical studies \cite{clastraj,cpl89,prl99,jhb33,pra63,pra66,pra65}, both classical trajectories analysis
and quantum mechanical computations, of inelastic cross sections and rate coefficients for collisions between
one atom and a rotationally excited diatom at ordinary and ultracold temperatures. The good qualitative agreement
between classical and quantum results suggested that the classical dynamics of the system is playing a significant role in 
the mechanism underlying the QRVR energy transfer.

Classical trajectory studies \cite{clastraj,cpl89,prl99} have shed light on the dynamics involved in this process.
It has been observed \cite{clastraj} that collisional processes that occur at high molecular rotation 
states and low collision velocities are composed of a sequence of small collisions, or {\sl collisionettes}, 
in which the interaction potential presents remarkable peaks. These {\sl collisionettes} take place whenever
the rapidly rotating diatom and the external atom come closer to each other, with the diatom in its outer vibrational 
turning point and the molecular axis nearly collinear with the atom. In between {\sl collisionettes} the interaction 
potential becomes insignificant and the rotational and vibrational actions remain nearly constant. 

The ratio between the change in the two molecular actions is approximately (and often spectacularly close to)
 a low order rational ratio of the vibrational and rotational classical molecular frequencies. This suggests that the vibrational and rotational frequencies of the diatom are in approximate low-order resonance
when that QRVR transfer occurs, and consequently a commensurate relation of the type
\beq\label{quas2}
\frac{p}{q}=\frac{n_v}{n_j} \simeq \frac{\omega_v}{\omega_j}
\eeq
holds, with  $n_v$ and $n_j$ small integers and $\omega_v$ and $\omega_j$ the classical vibrational
and rotational molecular frequencies.

The observed frequency-locking relation (\ref{quas2}) suggested that the interaction potential is acting as a nonlinear external
coupling between the internal vibrational and rotational molecular oscillators. This conjecture was later supported by Hoving {\sl et al.} 
\cite{cpl89} who explained QRVR transfer in terms of adiabatic invariance and presented numerical evidence which showed that 
this phenomenon is associated with the appearance of very large isolated nonlinear resonances in the phase of the system.
Numerical results also indicated that the general rule (\ref{quas1}) followed by the molecular rotational and vibrational quantum 
numbers in the dominant QRVR transitions, which according to (\ref{quas2}) can be expressed as
\beq\label{quas3}
n_{v}\Delta v + n_{j}\Delta v =0
\eeq
is connected to the conservation of the action
\beq\label{quas4}
I = n_v v + n_j j
\eeq

Quasiresonant energy transfer processes have also been observed in atom-diatom collisions that occur at ultralow 
energies \cite{prl99}. However,  in this regime there are new aspects that introduce qualitative differences between 
the classical and quantum behaviors. The strong correlation between $\Delta j$ and $\Delta v$ persists, but the
magnitude can be much smaller than one quantum, making the quantum transitions classically forbidden.
In the quantum limit,  significant threshold effects appear in the quenching rate coefficient in the zero temperature limit causing 
some of the most relevant quasiresonant channels to remain closed due to energy conservation.
Classically, the collisionettes described in processes at higher energies are replaced by a strong modulation of
the interaction potential at the characteristic frequency of the quasiresonant transition. For certain initial
conditions the incoming atom can get temporarily trapped around the diatom, bouncing several times before being
rejected by the repulsive component of the interaction potential. Although the temporal evolution of the atom
along these trapping trajectories is a extremely sensitive function of the initial conditions ({\sl chaotic
scattering region}), remarkably the quasiresonant correlation (\ref{quas1}) is still satisfied.

In this work we will focus on the classical analysis of  quasiresonance. We will show that this is a
common effect, not restricted to atom-diatom inelastic collisions, which arises from transient
"turning on" of interactions between {\sl quasi}-resonant, i.e. not necessarily exactly resonant, degrees of freedom.
The same quasiresonance effect will be shown to arise from an explicitly time dependent transient ``turning on" 
and one that occurs autonomously.

Although the work presented bears some relation with  {\sl adiabatic switching} \cite{switch}, the connection 
has not been explored. Adiabatic switching is a method for semiclassical quantization which had some success in starting with a 
multidimensional separable problem of known quantized actions and switching on interactions adiabatically, hoping that 
the actions do not change, and thus allowing the energy with those actions to be read off of the trajectories which have 
suffered the switching.

The paper is organized as follows. We begin in section \ref{adia} with the theoretical analysis of classical system 
under a transient interaction. In section \ref{num} we illustrate the main features of the quasiresonance effect numerically,
using  a two-oscillator resonant Hamiltonian, a new example of 
quasiresonant system. Section \ref{mechanism}  further analyzes the mechanisms behind quasiresonance, especially from the point of view of phase space. Here we also present a more detailed discussion of the quasiresonance effect between the vibrational and rotational
internal degrees of freedom of a diatom molecule in slow inelastic collisions with an external atom.
Finally, in section \ref{concl}  we summarize the  main conclusions of the work.

\section{Adiabatic invariants under a slow transient interaction}
\label{adia}
In this section we show that quasiresonance is a common effect which arises from the coupling of approximately 
resonant internal degrees of freedom of a classical system, perturbed by some transient interaction with additional degrees 
of freedom.
We start by considering a two-dimensional integrable system $H_0$,  perturbed by the transient interaction with one
structureless particle. The Hamiltonian of the total system may be expressed as  
\beq\label{tran1}
H=H_0({\bf J})+H_k({\bf P})+\epsilon V({\bf J},\uPfi,{\bf P},{\bf Q};\uxi)
\eeq
where ${\bf J}\equiv(J_1,J_2)$ and $\uPfi\equiv(\Phi_1,\Phi_2)$ are the action-angle variables that
describe the unperturbed system $H_0$, ${\bf Q}$ and ${\bf P}$ the coordinate and the conjugate momentum 
(not necessarily action-angle variables) of the incoming particle and $H_k$ its kinetic energy. 
$\epsilon$ is a small parameter which characterizes the magnitude of the interaction term. We will assume that the term satisfies
 the condition
\beq\label{tran2}
\lim_{\uxi\rightarrow\infty}V({\bf J},\uPfi,{\bf Q};\uxi)=0
\eeq
with $\uxi\equiv\uxi({\bf J},\uPfi,{\bf P},{\bf Q})$ an {\sl interaction parameter} that controls the amplitude of the
interaction between the system $H_{0}$ and the particle. In a typical collisional process this parameter would correspond 
to the distance between centers of mass of the colliding species, which do not notice the interaction potential unless they are close  enough to each other. 

\begin{figure}
\centering
\vspace*{0.65cm}
\includegraphics[height=5cm]{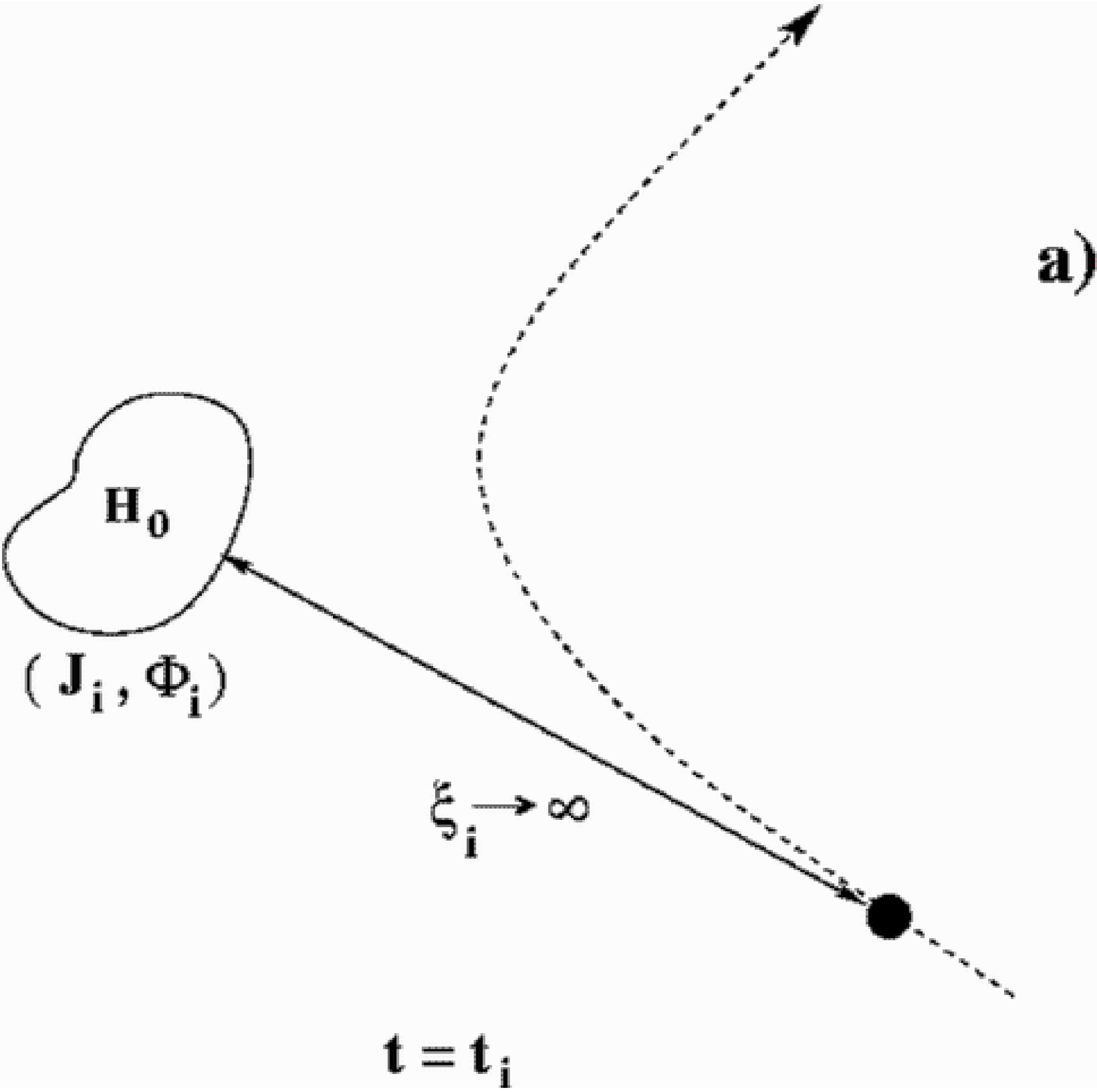}
\hspace*{2.9cm}
\includegraphics[height=5cm]{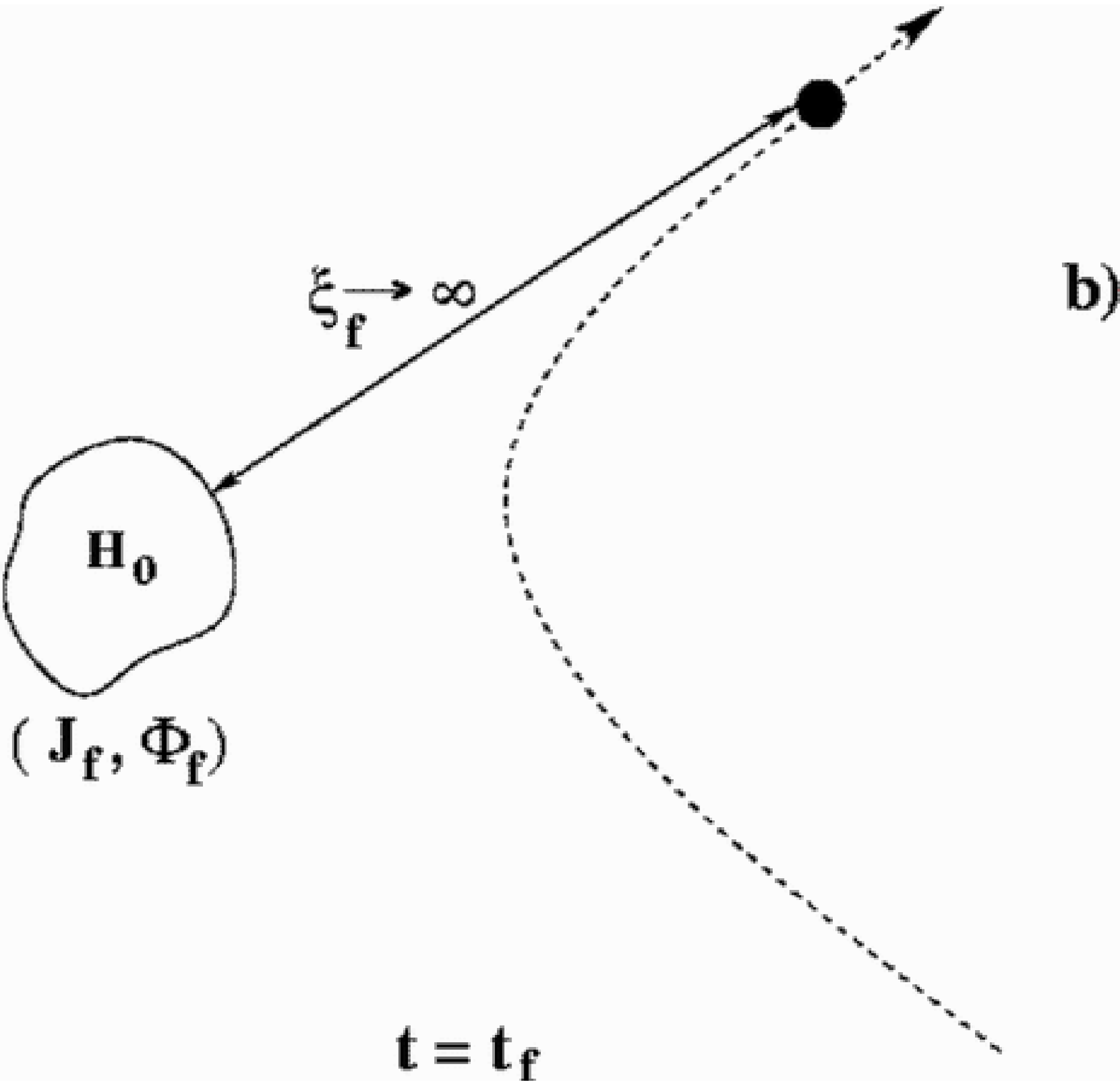}
\caption{\small\
A sketch of the initial $(a)$ and final $(b)$ states in the transient interaction process between the 
system $H_0$ a structureless incoming particle.
\label{fig_pict2}}
\end{figure}

We are interested in the analysis of a {\sl collisional event} in which the transient interaction of the system $H_{0}$
with the particle induces a change in its internal state from  the initial state 
$({\bf J_i},\uPfi_{\bf i})\equiv(J_{1i},J_{2i},\Phi_{1i},\Phi_{2i})$ at $t=t_i\,(\uxi_{i}\rightarrow\infty)$ 
to the final state $({\bf J_f},\uPfi_{\bf f})\equiv(J_{1f},J_{2f},\Phi_{1f}\Phi_{2f})$ at 
$t=t_f\,(\uxi_{f}\rightarrow\infty)$, see figure (\ref{fig_pict2}).

Let us assume that the two independent frequencies of the unperturbed Hamiltonian $H_0$ satisfy an approximate 
resonant condition of the form
\beq\label{tran3}
M\omega_2-N\omega_1\simeq 0
\eeq
where
\beq\label{tran4}
\omega_i({\bf J})=\frac{\partial H_0}{\partial J_i}
\hspace*{1cm}(i=1,2)
\eeq
The secularity in the unperturbed Hamiltonian due to this resonance condition can be removed applying standard
secular perturbation theory \cite{lich_lieb}. Namely, we can perform a canonical transformation from the variables 
$({\bf J},\uPfi)$ to a new set of action-angle variables $({\bf I},\uphi)$, which define a frame  that rotates with 
the resonant frequency. Taking the generating function for the canonical transformation as
\beq\label{tran4}
F({\bf I},{\bf P},\uPfi,{\bf Q})=(M\Phi_2-N\Phi_1)I_1+\Phi_2 I_2+{\bf P}\cdot{\bf Q}
\eeq
the equations for the transformation between the two set of variables of the Hamiltonian $H_0$ are given by
\beq\label{aver2}
\phi_1=\frac{\partial F}{\partial I_1}=M\Phi_2-N\Phi_1
\eeq
\beq\label{aver3}
\phi_2=\frac{\partial F}{\partial I_2}=\Phi_2
\eeq
\beq\label{aver4}
J_1=\frac{\partial F}{\partial \Phi_1}=-NI_1
\eeq
\beq\label{aver5}
J_2=\frac{\partial F}{\partial \Phi_2}=MI_1+I_2
\eeq
In terms of the new variables the transformed Hamiltonian may be expressed as
\beq\label{tran7a}
{H}'=H_0({\bf I})+H_k({\bf P})+\epsilon{V}({\bf I},\uphi,{\bf P},{\bf Q};\uxi)
\eeq
If the approximate resonance condition (\ref{tran3}) is satisfied, the evolution of the new angle variables in the 
rotating frame is given by 
\beq\label{tran7b}
{\dot\phi}_{1}=M{\dot\Phi}_{2}-N{\dot\Phi}_{1}\simeq\epsilon
\left(M\frac{\partial V}{\partial J_{2}}-N\frac{\partial V}{\partial J_{1}}\right)
\eeq
and
\beq\label{tran7c}
{\dot\phi}_{2}={\dot\Phi}_{2}=\omega_{2}({\bf I})+\epsilon
\frac{\partial V}{\partial J_{2}}
\eeq
Hence, provided that the frequency $\omega_{2}$ is far enough from zero, the oscillation of the new angle variable
$\phi_{2}$ near  the resonance will be fast compared to the variation of the angle variable $\phi_{1}$. Under these
conditions, the dynamics  of the system in the proximity of the nonlinear resonance can be described by the averaged 
Hamiltonian
\beq\label{tran7}
{\bar H}=H_0({\bf I})+H_k({\bf P})+\epsilon{\bar V}({\bf I},\phi_1,{\bf P},{\bf Q};\uxi)
\eeq
where
\beq\label{tran8}
{\bar V}({\bf I},\phi_1,{\bf P},{\bf Q};\uxi)=\frac{1}{2\pi}\int_0^{2\pi}
V({\bf I},\uphi,{\bf P},{\bf Q};\uxi)d\phi_{2}
\eeq
Since ${\bar H}$ does not depend on the angle variable $\phi_2$ its conjugated action variable $I_2$ is an adibatic invariant 
of the motion in the proximity of the resonance. But, according to (\ref{aver4}) and (\ref{aver5}), this adiabatic invariance 
of the action conjugated to the fast angle variable $\phi_2$ implies that
\beq\label{tran9}
I_2=J_2+\frac{M}{N}J_1=const
\eeq
or
\beq\label{tran10}
N\Delta J_2+M \Delta J_1=N\left(J_{2f}-J_{2i}\right)+M\left(J_{1f}-J_{1i}\right)=0
\eeq
which gives the rational ratio in the action changes that is observed in the quasiresonance effect.
Thus, a quasiresonance is implied by  the existence of an adiabatic invariant characterizing the 
quasiperiodic motion of the system, in the proximity of a nonlinear resonance between its internal degrees of freedom. 

From this perspective quasiresonance is merely the survival of an adiabatic action, corresponding to a fast angle.  However,  subtleties regarding the boundaries of adiabaticity as probed by the strength and duration of the interaction which couples the unperturbed actions leave a rich subject to be explored.
\vskip .1in
\subsection{Example: Atom-vibrotor collision}
\vskip .1in
A nice illustration of quasiresonance effect can be found in the dynamics of an atom,  fixed on a spring and 
is rotating on a plane, which is slightly perturbed by the collision with a slow incoming particle. Before the interaction 
occurs the dynamics of the unperturbed rotating atom is described by the integrable Hamiltonian
\beq\label{spin1}
H_{0}=H_{0}(J_{r},J_{\theta})=\frac{p_{r}^{2}}{2m}+\frac{p_{\theta}^{2}}{2mr^{2}}+U(r)
\eeq
where $m$ is the atom mass, $p_{r}$ and $p_{\theta}$ are the momenta conjugated to the radial and angular 
variables  $r$ and $\theta$, $U(r)$ the internal {\sl elastic} interaction potential, and 
\beq\label{spin2}
J_{r}=\frac{1}{2\pi}\oint p_{r}dr=\frac{1}{2\pi}\oint\sqrt{2m\left[H_{0}-U(r)\right]
-\frac{p_{\theta}^{2}}{r^{2}}}dr
\eeq
and
\beq\label{spin3}
J_{\theta}=\frac{1}{2\pi}\oint p_{\theta}d\theta=p_{\theta}
\eeq
the vibrational and rotational action variables.

\begin{figure}
\centering
\vspace*{0.65cm}
\includegraphics[height=5cm]{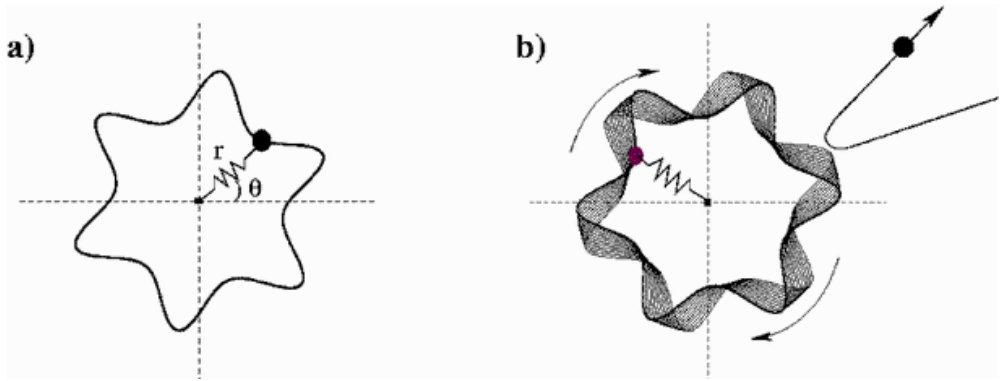}
\caption{\small\
In $a)$ the trajectory (a fixed ``gear" shape) described by an atom fixed on a spring when there is a integer ratio of six between 
the vibrational and rotational frequencies.  This is the form of the fast angle-averaged potential in the adiabatic analysis. In $b)$ is a sketch of  the evolution of the vib-rotor after it is perturbed by a slow 
colliding particle.
\label{fig_spin}}
\end{figure}

Initially the atom is vibrating in and out as it rotates on a plane, tracing out a ``gear" shape which is actually fixed in
space if the radial and angular frequencies satisfy a resonance condition (\ref{tran3}), see 
panel $a)$ in figure (\ref{fig_spin}).  More generally, the ``gear'' is slowly rotating clockwise or counterclockwise initially, with angular frequency ${\dot\phi}_{1}$, i.e. the slow angle. The gear-shaped potential is just the potential of the averaged Hamiltonian given in Eq.~\ref{tran7}.

The interaction with the incoming particle induces a coupling between the internal degrees 
of freedom of the spinning atom and, therefore, a change in its vibrational and rotational frequencies. In the proximity
of a resonance, when the condition $M\omega_{r}-N\omega_{\theta}\simeq 0$ is satisfied, such variation produces a slow 
rotation of the {\sl gear}, which starts moving as an effective {\sl rigid rotor}, see panel $b)$ in figure (\ref{fig_spin}). 
The angular momentum associated with this rotating gear is given by the new action variable $I_{\theta}$, while the action
conjugated to the rapid angle variable in the rotating coordinate system, in this system the radial or vibrational action, is the modified 
adiabatic invariant that satisfies the condition (\ref{tran9}). Namely,
\beq\label{spin4} 
I_{r}=J_{r}+\frac{M}{N}J_{\theta}=
\frac{1}{2\pi}\oint\sqrt{2m\left[H_{0}-U(r)\right]-\frac{p_{\theta}^{2}}{r^{2}}}dr
+\frac{M}{N}p_{\theta}=const
\eeq
This example illustrates the physical nature of the reduced dimensionality system which results when the average over the fast variable is performed.  As long as the quasiresonance holds, the reduced dimensionality object acts consistently as if it did not possess the degrees of freedom removed by averaging. 

\vspace*{0.1cm}

\vskip .1in
\subsection{Nonautonomous systems}
\vskip .1in
The physical picture of  the ``gear'' as the averaged Hamiltonian strongly suggests that 
an external perturbation pushing non-reactively (nonautonomously) could play the role of the incident atom in the example above, causing the gear to change precession in the same way as did collision with the atom.  Certainly the canonical transformation to slow and fast angle variables is still relevant,  as that involves only the ``system''. 
Thus we consider an integrable system perturbed by an explicitly time dependent interaction,
{\sl i.e.} the {\sl interaction parameter} is  $\uxi=\uxi(t)$. In terms of the action-angle variables 
$({\bf J},\uPfi)$ of the unperturbed system,  the total Hamiltonian may be expressed as
\beq\label{trant1}
H=H_0({\bf J})+\epsilon V({\bf J},\uPfi;\uxi)
\eeq
where $\epsilon$ is a small parameter that characterizes the perturbation strength, which  we will assume that
satisfies the {\sl transient} condition
\beq\label{trant2}
\lim_{t\rightarrow\pm\infty}V({\bf I},\uPfi;\uxi(t))=0
\eeq
In order to remove the singularity associated with the resonance (\ref{tran3}) between the unperturbed frequencies (\ref{tran4}), we
again seek a canonical  transformation from the variables $({\bf J},\uPfi)$ to a new variables  $({\bf I},\uphi)$ in the rotating coordinate
system.  We choose for such transformation the generating function
\beq\label{trant3}
F({\bf I},\uPfi,\uxi)=(M\Phi_2-N\Phi_1)I_1+\Phi_2 I_2+f({\bf I},\uPfi,\uxi)
\eeq
where $f$ is a multiply periodic function of the angles that can be expanded as the Fourier series
\beq\label{trant3b}
  f({\bf I},\uPfi,\uxi)=\sum_{n_{1},n_{2}}a_{n_{1}n_{2}}({\bf I},\uPfi,\uxi)e^{i(n_{1}\Phi_{1}+n_{2}\Phi_{2})}
\eeq  
With such generating function, the equations of the transformation between the two sets of variables  are 
\beq\label{trant4}
\phi_1=\frac{\partial F}{\partial I_1}=M\Phi_2-N\Phi_1
+\frac{\partial f}{\partial I_{1}}
\eeq
\beq\label{trant5}
\phi_2=\frac{\partial F}{\partial I_2}=\Phi_2
+\frac{\partial f}{\partial I_{2}}
\eeq
\beq\label{trant6}
J_1=\frac{\partial F}{\partial \Phi_1}=-NI_1
+\frac{\partial f}{\partial \Phi_{1}}
\eeq
\beq\label{trant7}
J_2=\frac{\partial F}{\partial \Phi_2}=MI_1+I_2
+\frac{\partial f}{\partial \Phi_{2}}
\eeq
In terms of the new variables, the transformed Hamiltonian can be written as
\beq\label{trant8}
{H}'=H({\bf I};\uxi)+\frac{\partial f}{\partial t}=
H({\bf I};\uxi)+{\dot{\uxi}}\frac{\partial f}{\partial \uxi}
\eeq
where $\dot{\uxi}=d\uxi/dt$. In the proximity of the resonance, the evolution of the new angle variables is given by
\beq\label{trant9}
{\dot\phi}_{1}= M{\dot\Phi}_{2}-N{\dot\Phi}_{1}+{\dot{\uxi}}\frac{\partial^{2} f}
{\partial\uxi\partial I_{1}}\simeq \epsilon\left(M\frac{\partial V}{\partial J_{2}}-N\frac{\partial V}
{\partial J_{1}}\right)+
{\dot{\uxi}}\frac{\partial^{2} f}{\partial\uxi\partial I_{1}}
\eeq
and
\beq\label{trant10}
{\dot\phi}_{2}={\dot\Phi}_{2}= \omega_{2}({\bf I})+\epsilon\frac{\partial V}{\partial J_{2}}+
{\dot{\uxi}}\frac{\partial^{2} f}{\partial\uxi \partial I_{2}}
\eeq

Let us now assume that the transient interaction is a function with slow time variation and, therefore,  we can consider ${\dot{\uxi}}$ 
as a small parameter that characterizes the slow change in the interaction. Provided that the frequency $\omega_{2}$ is sufficiently far 
from zero, (\ref{trant9}) and (\ref{trant10}) indicate that the angle $\phi_{2}$ oscillates much faster than $\phi_{1}$ near the resonance. 
Under these conditions, the slow motion of the system in this region can be described by the averaged Hamiltonian
 \beq\label{trant8}
{\bar H}=H({\bf I};\uxi)+\frac{1}{2\pi}\int_{0}^{2\pi}{\dot{\uxi}}\frac{\partial f({\bf I},\uPfi,\uxi)}{\partial \uxi}d\phi_{2}
\eeq
Thus, the application of the averaging over the fast angle leads to its conjugated action as the first term of the series for the modified adibatic invariant
of the system near a nonlinear resonance. As in the autonomous system, such a modified invariant reflects the change that the transient perturbation induces 
in the topology of the phase space trajectories in the proximity of a nonlinear resonance zone. To zero order in the perturbation strength and slowness 
parameters, the propensity rule associated with a quasiresonance arises also in a non-autonomous transient interaction from the linear relation between the 
modified invariant and the action variables of the unperturbed system. 

\section{Numerical studies and further  analysis of  quasiresonance}
\label{num}

We have just seen that the  quasiresonance can arise from a explicitly time dependent  or an autonomous transient interaction. 
A well defined model system suitable for further study is
composed of two anharmonic Morse type oscillators that are transiently coupled by a non-autonomous external perturbation.
In terms of the action-angle variables of the two uncoupled oscillators, we express this model resonant Hamiltonian as
\beq\label{hamil}
H=H_0({\bf J})+V(\uPfi,t)
\eeq
with the Hamiltonian for the unperturbed system
\beq\label{hamil0}
H_0({\bf J})=a_1J_1+a_2J_2 - a_{11}J_1^2 - a_{22}J_2^2
\eeq
where $a_i,a_{ii} \ge 0\,\,(i=1,2)$, and the non-autonomous transient perturbation as a sum of resonant
coupling terms
\beq\label{poten}
V(\uPfi,t)=v_0g(t)\sum_{n_1=1}^m\sum_{n_2=1}^m \sin(n_2\Phi_2-n_1\Phi_1)
\eeq
with $n_2$, $n_1$ and $m$ integers and $v_0$ the coupling strength. We will assume that the coupling 
interaction is turned on and turned off by means of a time-dependent a Gaussian function
\beq\label{latt3}
g(t)=\exp\left[-(t-t_p)^2/2\sigma^2\right]
\eeq
characterized by a parameter $\sigma$.

\begin{figure}[h]
\centering
\vspace*{0.65cm}
\includegraphics[height=8.9cm]{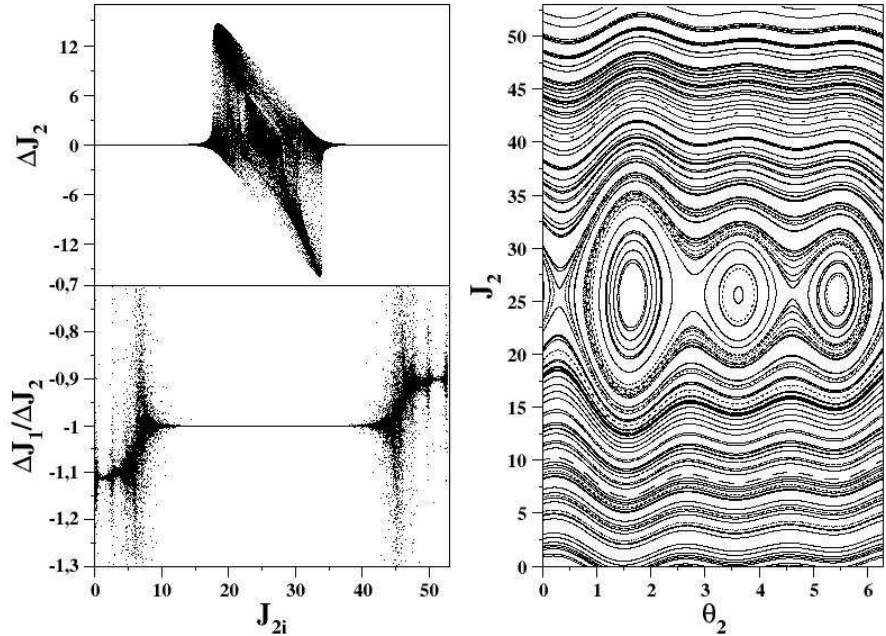}
\caption{\small\
On the left, the variation of the action $J_2$ (top) and the ratio of the changes in  the two actions,
$\Delta J_1/\Delta J_2$ (bottom), with respect to the initial value of the action $J_2$. Both pictures
include the data of $20$ classical trajectories with the same initial actions and angle variables
chosen at random. On the right, the $J_2-\theta_2$ phase plane corresponding to the trajectories included 
in the panels on the left. The time-dependent interaction potential was {\sl frozen} at time value $t=t_p$.
The total energy is $E=50$, the Gaussian parameter $\sigma=80$, the coupling strength
$v_0=0.05$ and $m=3$. The different parameters in the unperturbed Hamiltonian $H_0$ are $a_i=1$ and 
$a_{ii}=0.001\,\,(i=1,2)$. All quantities are expressed in arbitrary units.
\label{fig_sec1a}}
\end{figure}

 Figure (\ref{fig_sec1a}) displays the variation of the action $J_2$ with respect to its initial value
for trajectories with initial angle variables chosen at random.
We observe that the action remains nearly constant except in a well defined region of initial action values where 
significant changes occur. 
This region  corresponds  to values of the initial action where the quasiresonance
effect arises, with a low order rational value for the ratio of the changes in the actions of the unperturbed system.
Here the correlation between the variations of the actions of the two oscillators in the quasiresonance defines a 
large {\sl plateau} region characterized by the propensity rule
\beq\label{num1}
{\Delta J_1}+{\Delta J_2}=0
\eeq

Figure (\ref{fig_sec1a}) also shows a view of the $J_2-\theta_2$ phase plane when the coupling strength reaches its 
maximum amplitude at $t=t_p$. From the analysis of the previous  section it is clear that quasiresonance must be connected with the presence of  resonance zones in the phase
of the system \cite{cpl89}, and here we see them explicitly.  Although this surface of section was generated by examining the motion with the  coupling artificially fixed at the maximum value, later we shall examine the temporal evolution in phase space under the actual transient interaction.
As did Hoving and Parson \cite{cpl89} for  atom-diatom collisions, we find that the region of 
phase space corresponding to the values of the initial actions where a quasiresonance occurs is dominated by a strong isolated 
nonlinear resonance at fixed coupling. Here, we observe a large resonance zone composed of three islands which are centered at the initial action
value which gives the position of the center of the {\sl plateau} region defined by the propensity rule (\ref{num1}).
\vskip .1in
\subsection{ Resonance parallelogram}
\begin{figure}[h]
\centering
\vspace*{1.3cm}
\includegraphics[height=6.5cm]{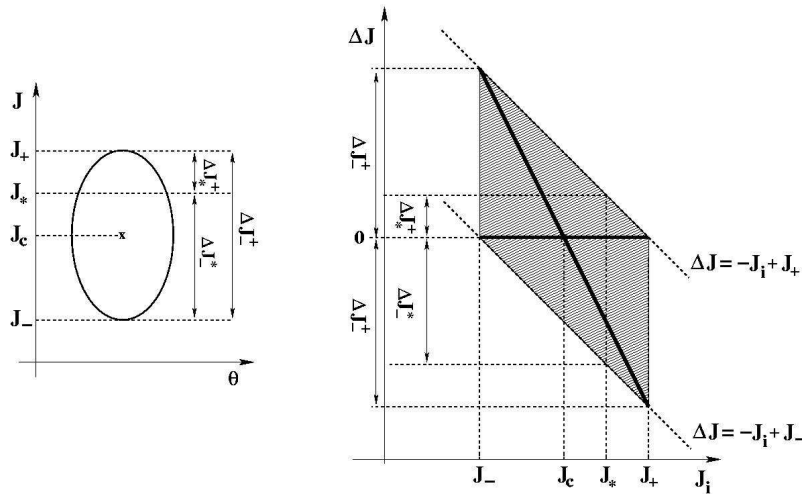}
\caption{\small\
On the left, a sketch of a resonance island in $J-\theta$ phase space at constant interaction strength. On the right, the shaded parallelogram 
represents the region that confines all the possible values of the change in the actions, $\Delta J=J_{f}-J_{i}$,  for the trajectories with initial 
actions $J_{i}$ within the limits of the resonance island on the left. The thick lines are the diagonals of the parallelogram, which are given by
$\Delta J=0$ and $\Delta J= -2J_{i}+2J_{c}$. 
\label{fig_Jchange}}
\end{figure}

\vskip .1in
Trajectories  can get caught inside the limits of a growing island and thereby
subject to action transport.   Thus, the parallelogram shape structure observed in figure (\ref{fig_sec1a}) for the change in the actions 
of trajectories included in a quasiresonance domain is  an image of a nearly uniform spread of the final action values within the limits of 
former resonance islands, see figure (\ref{fig_Jchange}).

\vskip .1in
\subsection{ How close is the action ratio to low order rationals?}
\vskip .1in
The hallmark  of quasiresonance is the ratio of action changes adhering to a low order rational, reflective of a near (but not necessarily exact) resonance in the system phase space. How close to the rational value are the action ratios in fact?  Linear plots are of no use, since the accuracy is typically far better than the linewidth of the plot. 
\begin{figure}[h]
\centering
\vspace*{1.3cm}
\includegraphics[height=8.9cm]{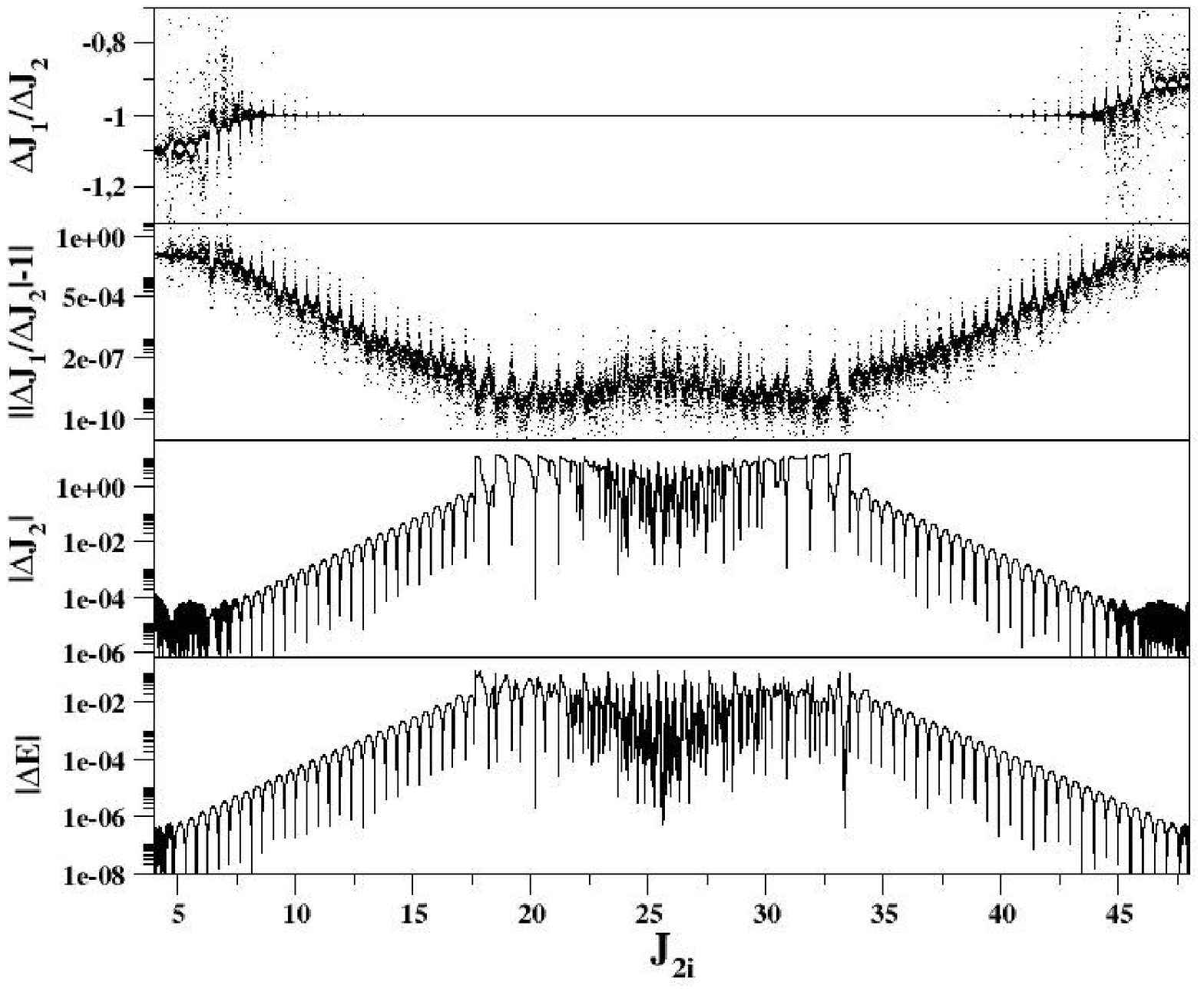}
\caption{\small\
From the top to the bottom, the ratio of action changes, the absolute value of the difference of the absolute value of this ratio 
from the rational number $r=1$, the absolute value of the change in the action variable $J_{2}$ and the absolute value of the 
change in the energy of the system {\sl vs} the initial action $J_{2i}$ for one of the trajectories included in figure (\ref{fig_sec1a}). 
\label{fig_plt}}
\end{figure}
 As figure (\ref{fig_plt}) indicates, the significant variations in the actions for the initial conditions
included in a plateau are correlated with relatively  large  changes  in the energy of the system. Naively, variations of
similar order of magnitude could be expected to occur in the ratio of action changes. 
However,  even when the resonant condition is only approximate,  the {\sl quasiresonant} condition $M\omega_2-N\omega_1\simeq 0$ may be satisfied to much higher accuracy.
The energy is also not conserved (in the non-autonomous systems) or it is exchanged between the perturber and the system (in the autonomous systems), but the action change ratio remains rational to a high degree of accuracy, often one part in $10^6$ for example in the present case.
As seen in figure (\ref{fig_plt}), the deviation of the ratio of action changes in a plateau from a low order 
rational value can be several orders of magnitude below the change in the energy 
or the individual actions.

At higher energy the phase space which is accesible to the system is larger and new regions of initial action values that satisfy 
different quasiresonance conditions may appear. 
\begin{figure}
\centering
\vspace*{0.6cm}
\includegraphics[height=11.9cm]{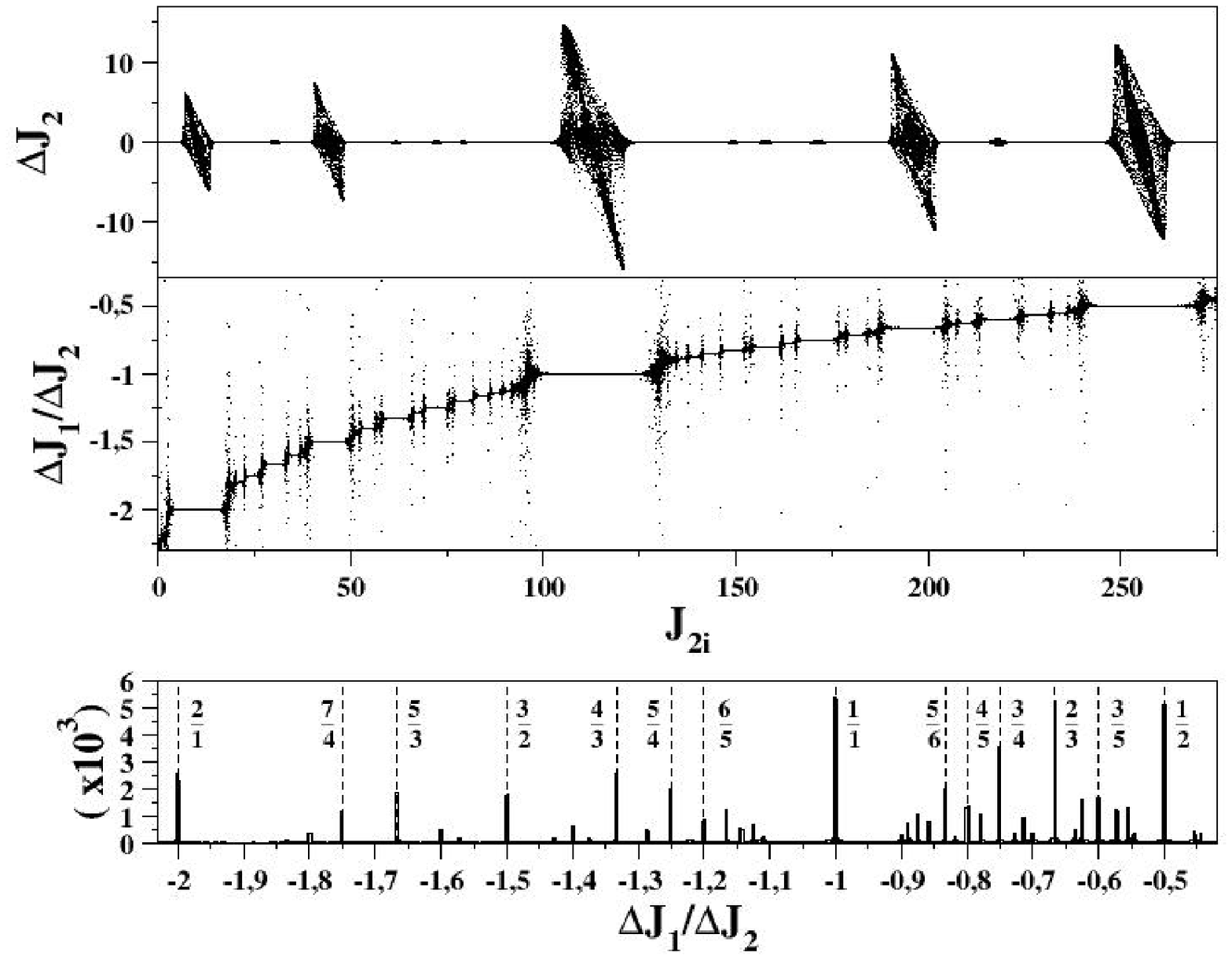}
\caption{\small\
The change in the action $J_{2}$ and the ratio of action changes {\sl vs} the initial action $J_{2i}$ for trajectories with the 
same parameters as in figure (\ref{fig_sec1a}), but energy $E=200$.
On the bottom, the histogram with the distribution of the values of the ratio of actions changes presented in the central panel.
The dashed lines indicate the locations of some low order rational values $r=n/m$ ($n$ and $m$ small integers), which coincide 
with the positions of the higher values in the histogram.
\label{fig_sec2a}}
\end{figure}
This is shown in figure (\ref{fig_sec2a}), where five regions that have significant changes in the actions can be seen. 
Each one of these regions defines a plateau in the ratio of action changes with a different rational value, $r=N/M=2/1,\,3/2,\,1/1,\,2/3$ 
and $1/2$,  in the propensity rule  (\ref{tran10}) that characterizes the quasiresonance.  
Besides these main plateaus, a staircase structure composed of  a series of small plateaus separated by regions of transition where 
the ratio of actions changes dramatically (quasiresonance is violated) is seen.  As figure (\ref{fig_sec2a}) also shows, these {\sl secondary plateaus}
are associated with quasiresonances characterized by large integer values $M$ and $N$ in the propensity rule (\ref{tran10}). 
Although in this figure the changes in the actions at these higher order quasiresonance domains are very small, and  some of them are 
hardly visible, we will see below that they become more important as the coupling between the internal degrees of freedom of the 
system increases, and that they are involved in the eventual demise  of the quasiresonance effect.

\begin{figure}
\centering
\vspace*{1.29cm}
\includegraphics[height=5.9cm]{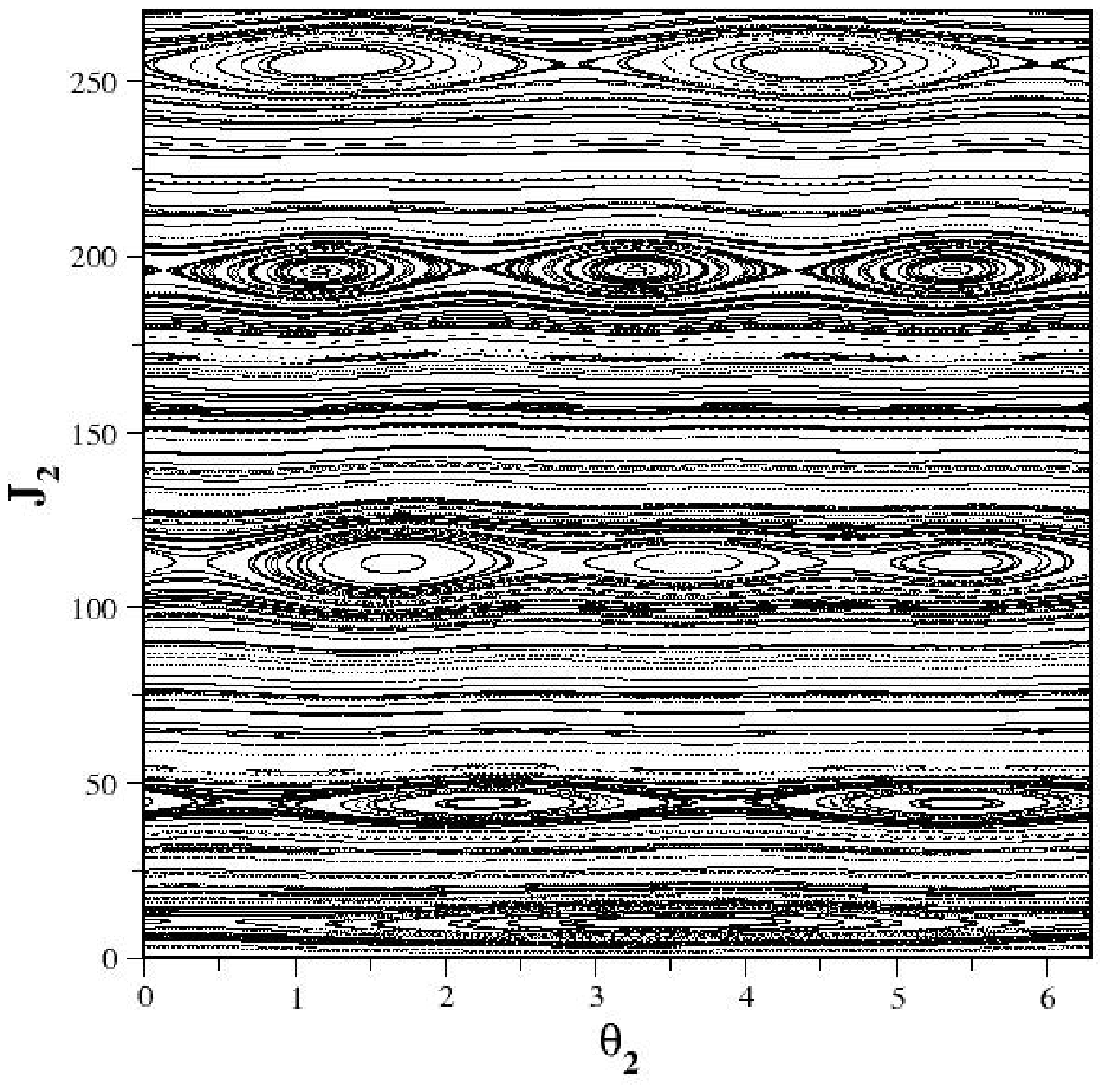}
\caption{\small\
The $J_2-\theta_2$ phase plane corresponding to the trajectories included in figure (\ref{fig_sec2a}).
The time-dependent transient interaction was {\sl frozen} at time value $t=t_p$.
\label{fig_sec1b}}
\end{figure}

As to be  expected \cite{cpl89}, the system with multiple quasiresonance plateaus now  presents a more complex phase space structure, see figure (\ref{fig_sec1b}),
Distinct resonance zones produce their own  quasiresonance domains. 
Immersed in the phase space regions between the main resonance zones it is possible to distinguish some long chains of tiny islands associated 
with the secondary plateaus that appear in figure (\ref{fig_sec2a}).
\vskip .1in
\subsection{ Interaction strength and quasiresonance}
\vskip .1in

In the perturbative regime, or weak coupling limit, when the energy of the unperturbed integrable system is much higher 
that the strength of the transient interaction, each main quasiresonance domain remains isolated. The corresponding plateaus in the ratio 
of action changes are well defined and are punctuated  by regions where the ratio of actions changes varies wildly. It should be noted that  in such non-quasiresonant regions  an approximate adiabatic invariance of
{\it both} actions of the unperturbed system holds, but the ratio of whatever small changes do occur is not locked at rational ratios.

As the strength of the coupling between the internal degrees of freedom of the perturbed system increases,  the
main quasiresonance domains grow, and higher order nonlinear resonances are activated in the regions between them. In the phase space, the generation of new quasiresonance domains is reflected by the
destruction of tori where the actions of the unperturbed system are adiabatic invariants and the emergence of chains small islands of regular 
motion associated with the modified adiabatic invariants in the proximity of higher order resonances.
The combined processes of generation of new higher order quasiresonances and the expansion of the existing ones make the edges of neighboring 
quasiresonance domains to get closer and finally come into contact. Afterwards, the overlap between different quasiresonance domains deteriorates 
them progressively and eventually is responsible for their disappearance. Hence, the Chirikov criterion \cite{Chirikov} on the overlapping of
neighboring resonances for the onset of chaos can be applied to estimate the destruction of a quasiresonance domain.
Once such overlap occurs, all the invariant tori between the island of the resonance zones are destroyed and regions of intermingled chaotic motion 
surrounding the islands of regular motion become visible.
In the strong coupling  limit, when all the resonance zones have came into contact, nearly all the phase space appears filled up by these stochastic regions
and only some isolated islands of the former main resonance zones may survive.
Under these conditions all regions of regular quasiperiodic motion shrink and therefore all adiabatic invariants cease to exist.   This picture of the demise of quasiresonance is approximate, however, since the timescale for the transient interaction can attenuate the effects of chaos; see Section~\ref{mechanism}. 

\begin{figure}
\centering
\vspace*{0.6cm}
\includegraphics[height=5.9cm]{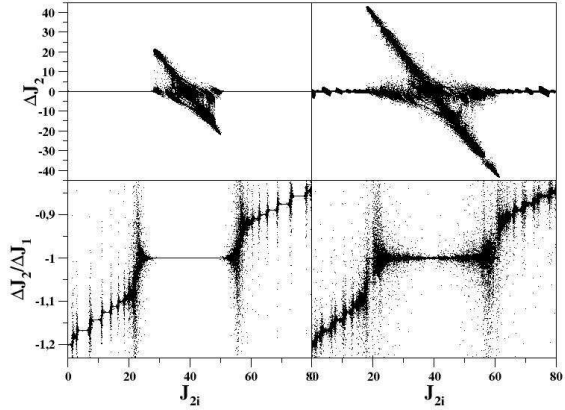}
\caption{\small\
The ratio of action changes and the change in the action $J_{2}$ {\sl vs} the initial action $J_{2i}$
for two different strengths  $v_{0}$ of the transient interaction. Panels on the left correspond to
$v_{0}=0.1$ and panels on the right to $v_{0}=0.4$. In both cases the energy is $E=75$, the Gaussian parameter 
in (\ref{latt3}) is $\sigma=100$ and up to $m=3$ coupling terms were included in the interaction term. The different 
parameters in the unperturbed Hamiltonian $H_0$ are $a_i=1$ and $a_{ii}=0.001\,\,(i=1,2)$. All quantities are 
expressed in arbitrary units.
\label{fig_v0a}}
\end{figure}

\begin{figure}
\centering
\vspace*{0.6cm}
\includegraphics[height=6.9cm]{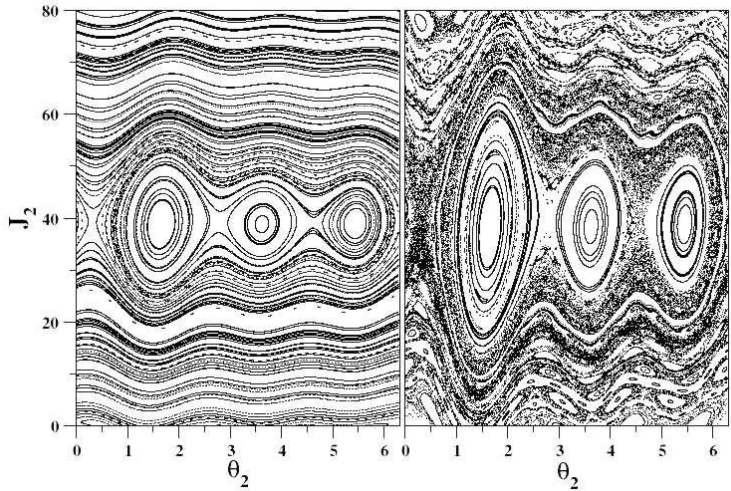}
\caption{\small\
The $J_2-\theta_2$ phase space for the trajectories included in figure (\ref{fig_v0a}) at $t=t_{p}$, when the
transient interaction reachs its maximum amplitude. The panel on the left corresponds to $v_{0}=0.1$ and the one 
on the right to $v_{0}=0.4$.
\label{fig_v0b}}
\end{figure}

In our model resonant Hamiltonian, the effects of a stronger coupling between the two oscillators on the quasiresonance can be 
analyzed taking larger amplitudes of the interaction strength $v_{0}$. As figures (\ref{fig_v0a}) and (\ref{fig_v0b}) show, the increase in  $v_{0}$ 
allows larger changes in the actions $J_{i}\,(i=1,2)$ around the resonant values. This causes an expansion of the resonances zones in phase space and
 makes more visible the long chains of small islands associated with higher order resonances. If the different resonance zones are well separated, an increase in the size of the corresponding quasiresonance domains might be expected. However, instead of that, figure 
 (\ref{fig_v0a}) shows is a deterioration of the  rational values for the ratio of action changes. The regions of stochastic behavior that emerge as a consequence of such overlap can be clearly observed in 
 figure (\ref{fig_v0b}). 

\begin{figure}
\centering
\vspace*{0.6cm}
\includegraphics[height=5.9cm]{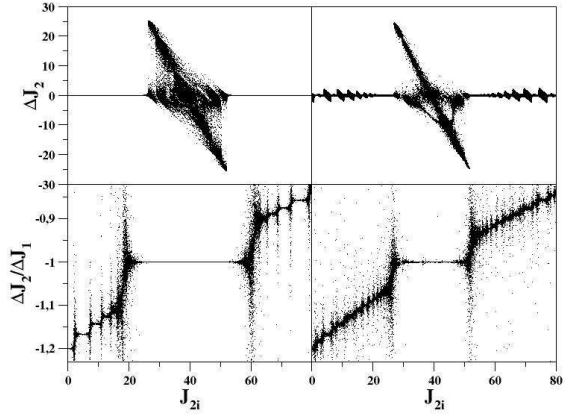}
\caption{\small\
The ratio of action changes and the change in the action $J_{2}$ {\sl vs} the initial action $J_{2i}$ for the {\sl normalized} transient interaction 
(\ref{norm3}). Panels on the left were obtained with $m=2$ and panels on the right with $m=5$. In both cases the energy is $E=75$,  the Gaussian 
parameter in (\ref{latt3}) is $\sigma=100$ and the coupling strength $v_{0}=0.4$. The parameters in the unperturbed Hamiltonian are the same as 
in the previous figures. All quantities are expressed in arbitrary units. 
\label{fig_ma}}
\end{figure}

\begin{figure}
\centering
\vspace*{0.6cm}
\includegraphics[height=6.9cm]{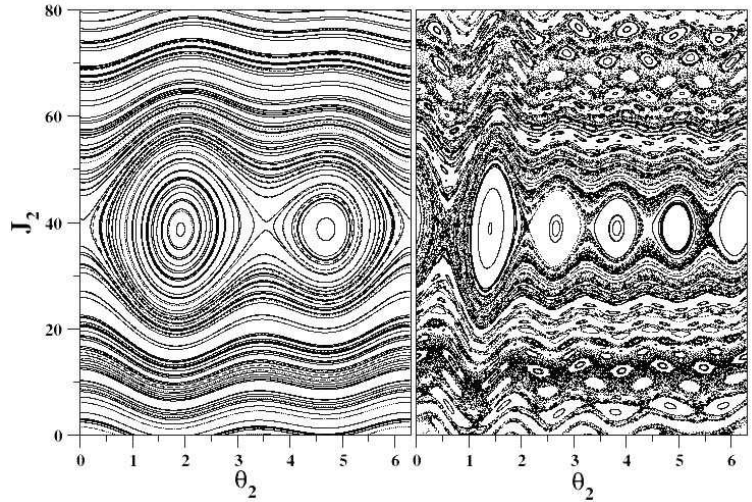}
\caption{\small\
The $J_{2}-\theta_{2}$ phase space for the trajectories included in figure (\ref{fig_ma}) at time $t=t_{p}$. The panel on the left corresponds 
to $m=2$ and the one on the right to $m=5$.
\label{fig_mb}}
\end{figure}

An alternative way to change the  coupling between the two oscillators s to include more coupling
terms in the transient interaction, while keeping constant its {\sl norm} in the angle variables. In this way we avoid the rapid expansion 
of the main resonance zones  and  introduce reduce the destruction 
of the main quasiresonance domains. In figures (\ref{fig_ma}) and (\ref{fig_mb}), for example, we set the 
norm as
\beq\label{norm1}
N_{1}=\frac{N_{m}}{m}=\sqrt{2}\pi,
\eeq
where
\beq\label{norm2}
N_{m}=\sqrt{\int_{0}^{2\pi}\int_{0}^{2\pi}\left[\sum_{n_1=1}^m\sum_{n_2=1}^m \sin(n_2\Phi_2-n_1\Phi_1)\right]^{2}
d\Phi_{1}d\Phi_{2}},
\eeq
by taking a {\sl normalized} interaction term given by 
\beq\label{norm3}
V(\uPfi,t)=\frac{v_0}{m}g(t)\sum_{n_1=1}^m\sum_{n_2=1}^m \sin(n_2\Phi_2-n_1\Phi_1)
\eeq
As figure (\ref{fig_ma}) shows, the increase in the number of coupling terms modifies the distribution of the final action values 
inside the main quasiresonance domains and induces higher order resonances, which give new secondary quasiresonance domains. 
Nonetheless, the emergence of higher order quasiresonances and the stronger overlap between neighboring domains eventually cause their destruction. 
The phase space of the system presents clear manifestations of these effects, see figure (\ref{fig_mb}), with an increase in the number of islands inside the 
main resonance zones and the proliferation of chains of small islands associated with high-order resonances. Also thick stochastic layers arising from the 
overlap of neighboring resonance zones become clearly visible.

\vskip .1in


\section{Phase space mechanism of Quasiresonance}
\label{mechanism}

\subsection{Energy dependence of the  resonance condition.}

The analysis of the phase space of the system has already shown the correspondence between the location
of the center of a {\sl plateau}  in the ratio of action changes in a quasiresonance and the center 
of a nonlinear resonance zone:  the center of the {\sl plateau} corresponding to the propensity rule (\ref{tran10}) coincides
with the location of the rational unperturbed torus whose independent frequencies satisfy a resonant
condition of the form (\ref{tran3}).
From now on we will denote as $J_{i,E}^{N:M}\,(i=1,2)$ the values of the actions of the unperturbed
system that give the location of this resonant torus at energy $E$. That is, the solutions of the
energy equation
\beq\label{frec0}
E=H_0({\bf J})
\eeq
and the resonant condition (\ref{tran3}). 

\begin{figure}
\centering
\vspace*{1.3cm}
\includegraphics[height=6.9cm]{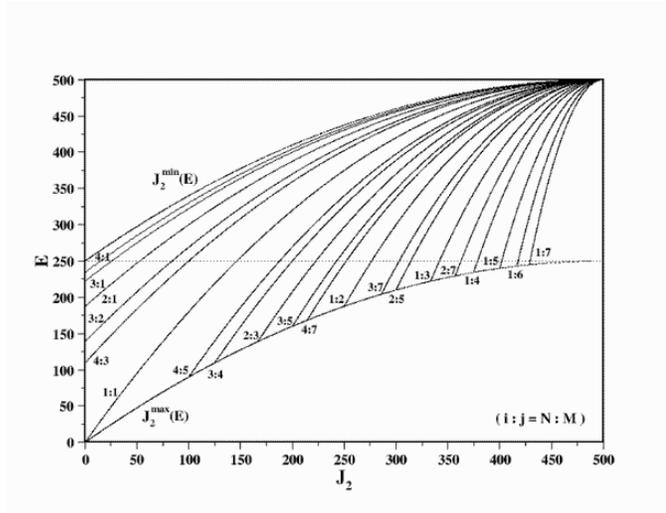}
\caption{\small\
Location in the $E-J_2$ plane of the lowest-order $N:M$ resonant tori of unperturbed the Hamiltoninan (\ref{hamil0}). 
The values taken for the parameters of the model are $a_i=1$ and $a_{ii}=0.001\,\,(i=1,2)$. All quantities are expressed
 in arbitrary units.
\label{fig_reso}}
\end{figure}

In the model resonant Hamiltonian introduced in the previous section, for example,
the unperturbed Hamiltonian $H_0$ defines an integrable system in which the actions of the two
uncoupled oscillators, $J_i\,(i=1,2)$, are conserved and the angle variables vary linearly in time,
$\Phi_i=\omega_i({\bf J})t+\Phi_{i0}$, with $\omega_i$ the unperturbed frequencies
\beq\label{w1}
\omega_i({\bf J})=\frac{\partial H_0}{\partial J_i}=a_i-2a_{ii} J_i
\hspace*{1cm}(i=1,2)
\eeq
The four-dimensional phase space of the unperturbed system is completely
stratified into two-dimensional tori defined by angle coordinates $(\Phi_1,\Phi_2)$ 
and radii $(J_1,J_2)$, with all periodic orbits restricted to lie on those invariant tori 
whose independent frequencies satisfy a commensurate relation. Thus, according to (\ref{hamil0}) and (\ref{tran3}), 
the actions on the rational unperturbed torus $N:M$ are related by:
\beq\label{reso2}
J_{2,E}^{N:M}=\frac{a_2}{2a_{22}}-\frac{1}{2a_{22}}\left(\frac{N}{M}\right)(a_1-2a_{11} J_{1,E}^{N:M})
\eeq
Figure (\ref{fig_reso}) shows the location in the $E-J_2$ plane of the lowest-order unperturbed
resonant tori in the range in which the unperturbed frequencies are positive,
$0 \le J_i \le J_{i}^{max}=\frac{a_i}{2a_{ii}}\,\,(i=1,2)$ and
$0 \le E \le E_{max}=\frac{1}{4} \left(\frac{a_1^2}{a_{11}}+\frac{a_2^2}{a_{22}}\right)$.

As soon as the transient interaction is applied the unperturbed tori of the system
begin to be deformed, but as we know from the Kolmogorov-Arnol'd-Moser (KAM) theorem,
not all of them in the same way. For sufficiently small coupling strength and Jacobians
$\partial{\uomega}/\partial{\bf J}$ different from zero, this theorem guarantees that
those tori bearing conditionally periodic motion with incommesurate frequencies
continue to exist, being only slightly distorted. Rational tori  whose independent frequencies satisfy a commensurate relation
(\ref{tran3}) are grossly deformed or even completely destroyed.

\subsection{ Span of a quasiresonance domain}

We  analyze  the phase space structures associated with the frozen, or constant transient 
interaction strength limit of the Hamiltonian. Although the interaction strength is always changing in quasiresonance,  fixed strength phase space plots are still useful, in particular for estimating the span of a quasiresonance domain.

We use our model resonant Hamiltonian (\ref{hamil}). It is convenient for our purposes to introduce dimensionless variables. 
Considering the scale of time as
\beq\label{esc1}
\tau=\frac{t}{\sigma}
\eeq
and introducing dimensionless action variables
\beq\label{esc2}
j_i=\frac{J_i}{\sigma E} \hspace*{1cm}(i=1,\,2)
\eeq
with $E$ the initial energy of the unperturbed system, we write the scaled Hamiltonian
\beq\label{esc3}
h=\frac{H}{E}
\eeq
as
\beq\label{esc4}
h=s_1j_1+s_2j_2-s_{11}j_1^2-s_{22}j_2^2+
\varepsilon g(\tau)\sum_{n_1=1}^m\sum_{n_2=1}^m\sin(n_2\Phi_2-n_1\Phi_1)
\eeq
where $s_i=\sigma a_i$, $s_{ii}=E\sigma^2a_{ii}\,(i=1,2)$, $\varepsilon=v_0/E$ and 
\beq\label{esc5}
g(\tau)=\exp\left[-(\tau-\tau_p)^2/2\right],
\eeq
with $\tau_p=t_p/\sigma$.
We will assume below that the total energy of the unperturbed system is much higher than the maximum coupling 
strength, $\varepsilon<<1$. In this limit the analysis of the dynamics of the system using first order classical 
perturbation theory will provide a good approach to describe the coupling process between the two oscillators.

\begin{figure}
\centering
\vspace*{0.6cm}
\includegraphics[height=6.9cm]{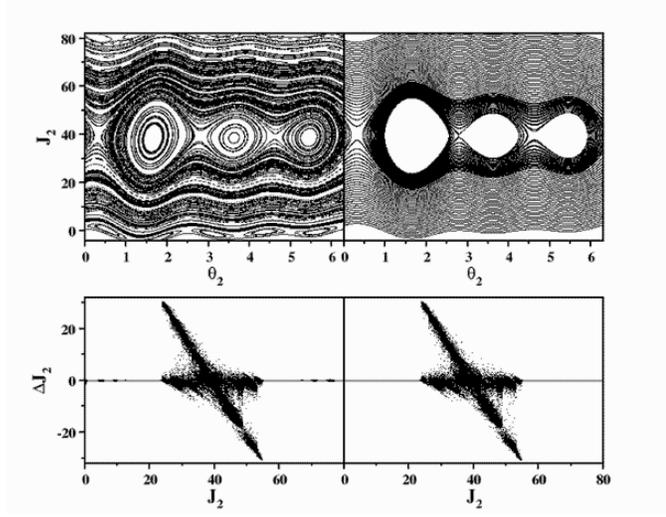}
\caption{\small\
Poincare surfaces of section for the total Hamiltonian (\ref{esc4}) (on the left top) and for the 
averaged Hamiltonian (\ref{aver7a}) (on the right top) at time $t=t_p$, total energy $E=75$,
Gaussian parameter $\sigma=100$, coupling strength $v_0=0.2$ and $m=3$. On the bottom the 
corresponding variations of the action $J_2$.
\label{fig_sec1}}
\end{figure}

Let us now assume that the two frequencies of the unperturbed oscillators satisfy the resonant condition (\ref{tran3}). 
The removal of the resonant variables applying the canonical transformation given by the generating function
\beq\label{aver1}
F=(M\Phi_2-N\Phi_1)I_1+\Phi_2 I_2
\eeq
leads to the Hamiltonian
\beqn\label{aver6}
h=(s_2 M-s_1 N)I_1+s_2 I_2-(s_{11} N^2 + s_{22} M^2)I_1^2-2 s_{22} MI_1I_2-s_{22} I_2^2 + \\ \nonumber
+\varepsilon g(\tau) \sum_{n_1=1}^{m}\sum_{n_2=1}^{m}
\sin\left\{\left(n_2-n_1\frac{M}{N}\right)\phi_2+\frac{n_1}{N}\phi_1\right\}
\eeqn
The averaging of this transformed Hamiltonian over the rapid oscillation of the new angle variable $\phi_2$
near the resonance zone gives the secular Hamiltonian
\beq\label{aver7a}
{\bar h}=(s_2 M- s_1 N)I_1+s_2 I_2-(s_{11} N^2+s_{22} M^2)I_1^2-2 s_{22} MI_1I_2-s_{22} I_2^2
+\varepsilon g(\tau)\sum_{k=1}^{m}\sin(k\phi_1)
\eeq
Thus the dynamics of the system in the proximity of the resonance $N:M$ can be described by the equations of motion
\beq\label{aver10a}
\frac{d\phi_1}{d\tau}=(s_2 M-s_1 N)-2(s_{11} N^2+s_{22} M^2)I_1-2 s_{22} MI_2
\eeq
\beq\label{aver11a}
\frac{dI_1}{d\tau}=-\varepsilon g(\tau)\sum_{k=1}^{m}k\cos(k\phi_1)
\eeq
Our previous numerical results show that the length of the plateau regions in the ratio of action changes 
can be estimated from the maximum size of the nonlinear resonance zones. 
As was expected, see figure (\ref{fig_sec1}), the phase space of the averaged Hamiltonian
reproduces the main islands of the resonance zone, but not  the chains of small islands associated
with higher-order resonances and neither any stochastic region. 

The fixed points of the secular Hamiltonian occur at the action value  
\beq\label{aver9}
I_{1}^{N:M}=\frac{(s_2 M- s_1 N)-2 s_{22} MI_2}{2(s_{11} N^2+s_{22} M^2)}
\eeq
which corresponds to the resonant actions $J_{i,E}^{N:M}\,(i=1,2)$ of the unperturbed Hamiltonian (\ref{hamil0}).
In the weak coupling limit, the dynamics of the system in the proximity of the resonance actions can be described by the 
{\sl pendulum} Hamiltonian
\beq\label{aver10}
{\bar h}_{pend}=-(s_{11} N^2+s_{22} M^2)(\Delta I_{1}^{N:M})^2+\varepsilon g(\tau)\sum_{k=1}^{m}\sin(k\phi_1)
\eeq
with $\Delta I_{1}^{N:M}=I_1-I_{1}^{N:M}$.

\begin{figure}
\centering
\vspace*{1.45cm}
\includegraphics[height=5.9cm]{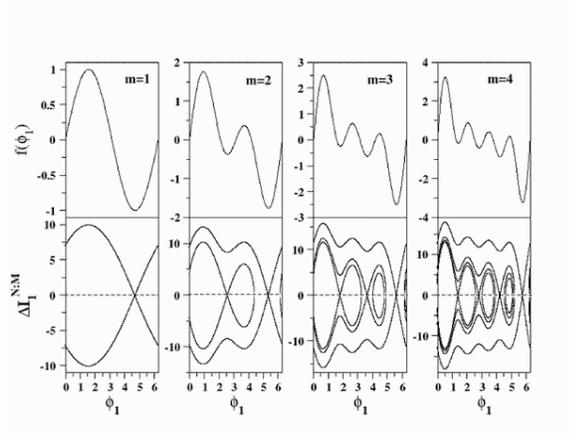}
\caption{\small\ The function  $f(\phi_1)$ (\ref{aver13}) and the curves 
of the separatrices $\Upsilon_k^{N:M}(\tau,\phi_1)\,(k=1,..,m)$ (\ref{aver14}) for the {\sl pendulum}
Hamiltonian ${\bar h}_{pend}$ (\ref{aver10}) for several values of $m$. The values taken for the different 
parameters of the model were; $N=M=1$, $a_{11}=a_{22}=0.001$, $v_0=0.1$ and $\tau=\tau_p$.
\label{fig_sep0}}
\end{figure}

The $2m$ fixed points of ${\bar h}_{pend}$ occur at the resonance action $I_1=I_{1}^{N:M}$, the $m$
elliptic fixed points, $0<\phi_{1,1}^s<\phi_{1,2}^s<..<\phi_{1,m}^s<2\pi$, correspond to maximum values of the function 
\beq\label{aver13}
f(\phi_1)=\sum_{k=1}^{m}\sin(k\phi_1)
\eeq
and $m$ the hyperbolic fixed points, $0<\phi_{1,1}^u<\phi_{1,2}^u<..<\phi_{1,m}^u<2\pi$, to its minimum values. 
The curve of the separatrix that passes through the hyperbolic fixed point $\phi_{1,k}^u\,(k=1,..,m)$
is given by:
\beq\label{aver14}
\Upsilon_k^{N:M}(\tau,\phi_1)=\sqrt{\frac{\varepsilon g(\tau)\left[f(\phi_1)-f(\phi_{1,k}^u)\right]}{(s_{11} 
N^2+s_{22} M^2)}}
\hspace*{0.6cm}(k=1,...,m)
\eeq
Figure (\ref{fig_sep0}) shows the function $f(\phi_1)$ and the curves of the separatrices 
$\Upsilon_k^{N:M}(\tau,\phi_1)$ for different values of $m$. 
The case $m=1$, with only one hyperbolic fixed point, corresponds to a 
single pendulum Hamiltonian. For $m>1$ there are an increasing number of hyperbolic fixed points which
develop a structure composed of $m$ {\sl encapsulated} separatrices curves. 
The outermost separatrix is associated with the deepest well in the function $f(\phi_{1})$ and 
passes through to the hyperbolic fixed point $\phi_{1,m}^u$. Whereas the innermost one is associated 
with the highest well and passes through the hyperbolic fixed point $\phi_{1,1}^u$.

The maximum excursion of the action $I_1$ along each separatrix curve $\Upsilon_k^{N:M}(\tau,\phi_1)$ occurs at 
the elliptic fixed point $\phi_{1,1}^s$. Thus the maximum variation of this action at each instant 
$\tau$ can be written as:
\beq\label{aver15}
\left[\Delta I_{1,k}^{N:M}(\tau)\right]_{max}=2\sqrt{\frac{\varepsilon g(\tau)\left[f(\phi_{1,1}^s)-
f(\phi_{1,k}^u)\right]}{(s_{11} N^2+s_{22} M^2)}}
\hspace*{0.6cm}(k=1,...,m)
\eeq

\begin{figure}
\centering
\vspace*{1.36cm}
\includegraphics[height=5.9cm]{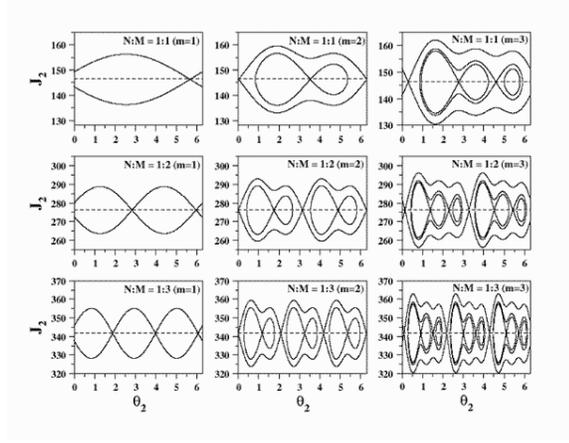}
\caption{\small\ Separatrices curves (\ref{aver14}) in the $J_2-\Phi_2$ phase plane.
The values taken for the different parameters of the model were: $a_{11}=a_{22}=0.001$, $v_0=0.1$ and time $\tau=\tau_p$.
The dashed line gives the location of the resonance action $J_{2,E}^{N:M}$ at energy $E=250$ (arb. units).
\label{fig_sep1}}
\end{figure}

Returning  to the initial action-angle variables to study the structure associated with the
resonance $N:M$ in the $j_1-\Phi_1\,(j_2-\Phi_2)$ phase plane, 
the center of the resonance zone is located at $j_1=j_{1,E}^{N:M}\,(j_2=j_{2,E}^{N:M}$).
According to the transformation relations (\ref{aver4})-(\ref{aver5}) and the equation (\ref{aver13})
there are a total number of $2mN\,(2mM)$ fixed points, $0<\Phi_{1,1}^s<\Phi_{1,1}^u<
\Phi_{1,2}^s<\Phi_{1,2}^u<...<\Phi_{1,mN}^s<\Phi_{1,mN}^u<2\pi\,
(0<\Phi_{2,1}^s<\Phi_{2,1}^u<\Phi_{2,2}^s<\Phi_{2,2}^u<...<\Phi_{2,mM}^s<\Phi_{2,mM}^u<2\pi)$.
Hence the curves of the separatrices associated with the hyperbolic fixed points create a {\sl chain} of 
$N\,(M)$ identical units, each of them containing $m$ {\sl encapsulated} separatrices curves. 
Figure (\ref{fig_sep1}) displays some of these chains.

The maximum deviation of  the action $j_1\,(j_2)$ takes from its resonant value $j_{1,E}^{N:M}\,(j_{2,E}^{N:M})$ 
along each separatrix curve is the same at all the elliptic fixed points $\Phi_{1,pm+1}^s\,(\forall p=0,..,N-1)$ 
$[\Phi_{2,pm+1}^s\,(\forall p=0,..,M-1)]$. 
Thus it follows from (\ref{aver4})-(\ref{aver5}) and (\ref{aver15}) that the maximum variation induced by the 
$N:M$ resonance zone on the actions $j_1$ and $j_2$ along the curve of the separatrix $k$ at time $\tau$ can be expressed as:
\beq\label{aver17}
\left[\Delta j_{1,k}^{N:M}(\tau)\right]_{max}=2N\sqrt{\frac{\varepsilon g(\tau)\left[f(\phi_{1,1}^s)-
f(\phi_{1,k}^u)\right]}{(s_{11} N^2+s_{22} M^2)}}
\hspace*{0.6cm}(k=1,...,m)
\eeq
and 
\beq\label{aver18}
\left[\Delta j_{2,k}^{N:M}(\tau)\right]_{max}=2M\sqrt{\frac{\varepsilon g(\tau)\left[f(\phi_{1,1}^s)-
f(\phi_{1,k}^u)\right]}{(s_{11} N^2+s_{22} M^2)}}
\hspace*{0.6cm}(k=1,...,m)
\eeq
We define  $J_{i,E,k}^{N:M\pm}\,(i=1,2)$ as the two extreme values of the action $J_i$ 
along the curve of the separatrix $k$ when the resonance zone 
reachs its maximum size at time $\tau=\tau_p$. That is,
\beq\label{aver19}
J_{i,E,k}^{N:M\pm}=J_{i,E}^{N:M}\pm\frac{1}{2}\left[\Delta J_{i,k}^{N:M}(\tau_p)\right]_{max}
\hspace*{1cm}(i=1,2\,\,;\,\,k=1,..,m)
\eeq

\begin{figure}
\centering
\vspace*{0.5cm}
\includegraphics[height=9.9cm]{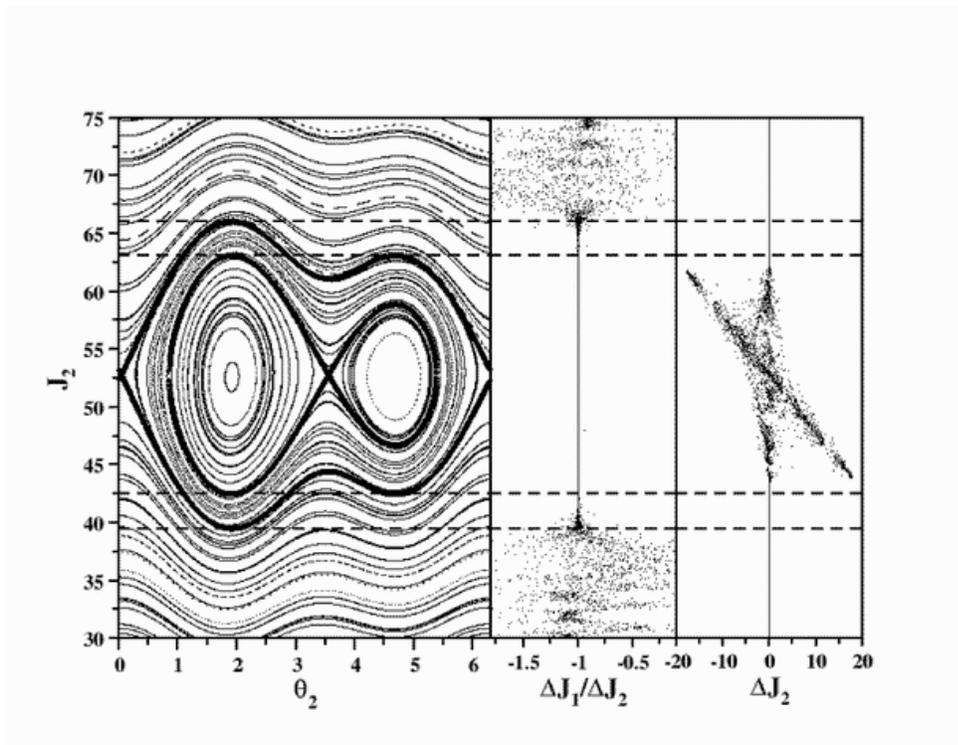}
\caption{\small\ 
On the left is shown the Poincare surface section in the $J_2-\Phi_2$ phase plane. The thick solid lines are the curves of the
separatrices (\ref{aver14}) for $N:M=1:1$, $k=1,2$ and $t=t_p$. The dashed lines give the values of the actions
$J_{i,E,k}^{N:M\pm}$ (\ref{aver19}) for the different separatrices.
In the central panel, the ratio of action changes, $\Delta J_1/\Delta J_2$, {\sl vs} the initial
action $J_2$. In the panel on the right the variation of the action $J_2$ with respect to its initial value.
In these last two pictures the data of $20$ trajectories with initial angles chosen at random are represented for each
value of the initial action $J_2$.
All data were obtained using the same parameter values: energy $E=100$, coupling strength $v_0=0.1$, temporal Gaussian
parameter $\sigma=200$, $m=2$, $a_1=a_2=1$ and $a_{11}=a_{22}=0.001$. All quantities are expressed in arbitrary units.
\label{fig_t1}}
\end{figure}
\begin{figure}
\centering
\includegraphics[height=8.9cm]{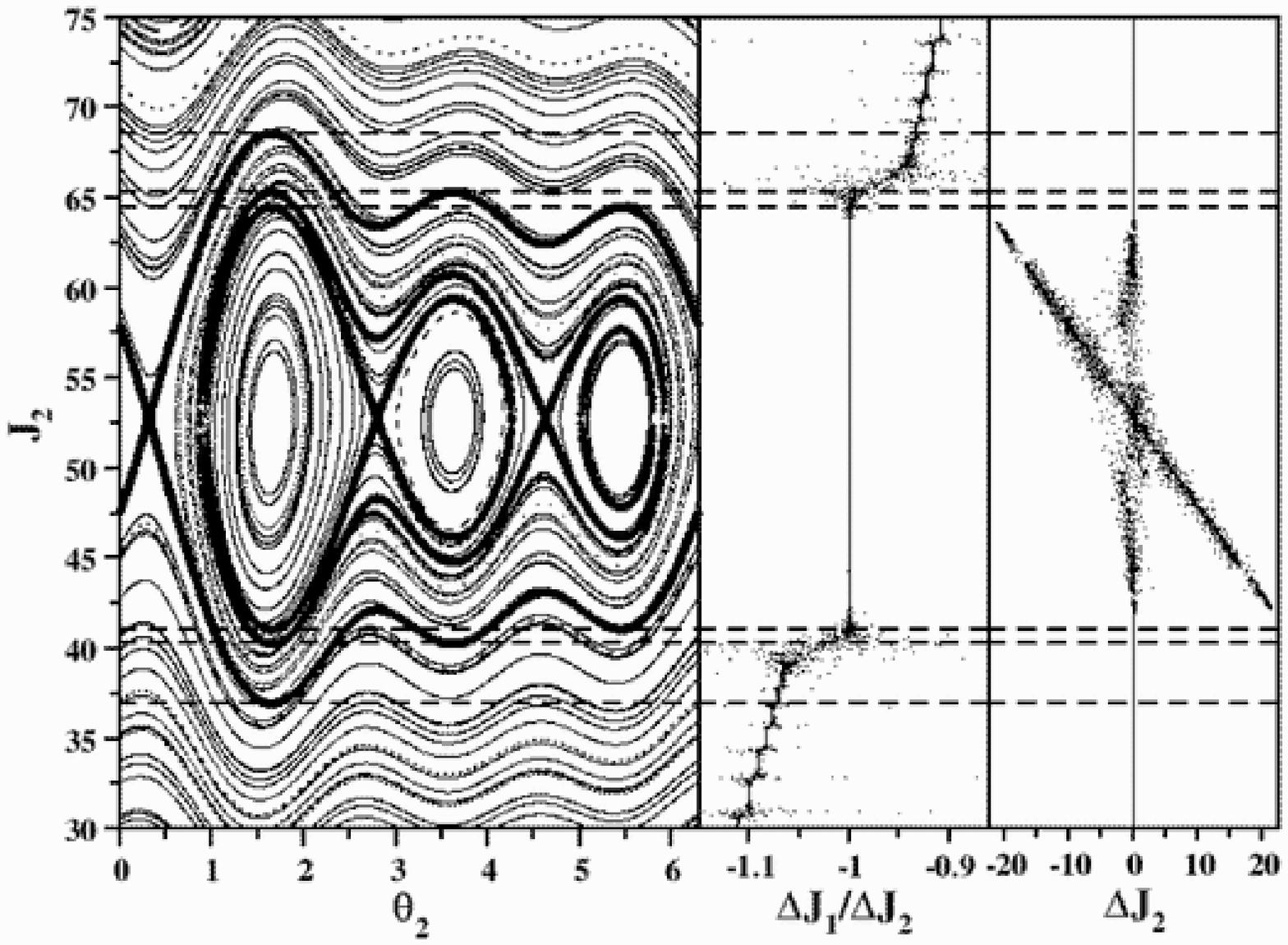}
\caption{\small\ 
The same as figure (\ref{fig_t1}) but for $m=3$.
\label{fig_t2}}
\end{figure}

We can now compare our analytical prediction for the size and shape of a primary resonance zone with the 
length of the corresponding plateau in the ratio of action changes in the quasiresonance effect.
Figures (\ref{fig_t1}) and (\ref{fig_t2}) display the quasiresonance domain  associated with the nonlinear
resonance $N:M=1:1$ for different number $m$ of coupling terms.
As was expected, the resonance zone presents a chain composed of $mM$ islands, two in figure (\ref{fig_t1}) 
and three in figure ({\ref{fig_t2}), associated with the $mM$ elliptic and $mM$ hyperbolic fixed points located 
at the resonance action $J_{2,E}^{1:1}$. 
The thick lines on the Poincare surface of sections represent the curves of the separatrices  
$\Upsilon_k^{N:M}(\tau_p,\phi_1)\,(k=1,...m)$ (\ref{aver14}) when the transient interaction reachs its maximum amplitude
at $\tau=\tau_p$. There is good agreement between the Poincare surface of sections obtained from the numerical simulations resonance analysis for the main resonance zones. Thus, in the limit of weak coupling between the two oscillators ($\varepsilon<<1$) the  expressions 
(\ref{aver17}) and (\ref{aver18}) give a good estimate of the maximum spread of the actions across the phase space region dominated
by the main resonance $N:M$.
Importantly, as figures (\ref{fig_t1}) and (\ref{fig_t2}) show, the maximum variation of the action along the curve of the 
innermost separatrix, $\left[\Delta J_{i,1}^{N:M}(t_p)\right]_{max}(i=1,2)$ (\ref{aver18}), provides a good approximation for the 
length of the plateau in the ratio of action changes in the quasiresonance. 

The phase space between the innermost and outermost separatrices corresponds  to deviations from the rational value of
the action changes giving the quasiresonance domain. This is due to the development of higher order resonance zones. The effect of these secondary resonances increases when the coupling between the 
two oscillators becomes stronger and eventually makes all the plateau region to disappear.

\vskip .1in
\subsection{ Effect of timescale of the transient interaction}
\vskip .1in
In the previous sections we have analyzed the phase space of the {\sl frozen} Hamiltonian. Now, 
in order to understand the behavior of the actions when a quasiresonance occurs, we focus on the time  evolution of the system 
as the external transient interaction is switched on and off.

The quasiresonance effect belongs to the larger class of adiabatic phenomena.  If the interaction turns on and off too fast,  adiabaticity will be violated. The timescale is set by the nature of the interaction potential and how often it is ``visited'' by the trajectory of the system, which in turn depends on the initial actions; see Fig.~\ref{newflower}. The period of  the visitation of the interaction potential must be much shorter than the  time scale of the transient perturbation. The interaction potential  may be localized to one or a few small regions in both angles, 
$(\Phi_1,\Phi_2)$, as it is in an atom-diatom collision for example. Or, it may be more extended, greatly affecting which resonance zones survive to cause quasiresonance plateaus.

\begin{figure}
\centering
\vspace*{0.6cm}
\includegraphics[height=6.9cm]{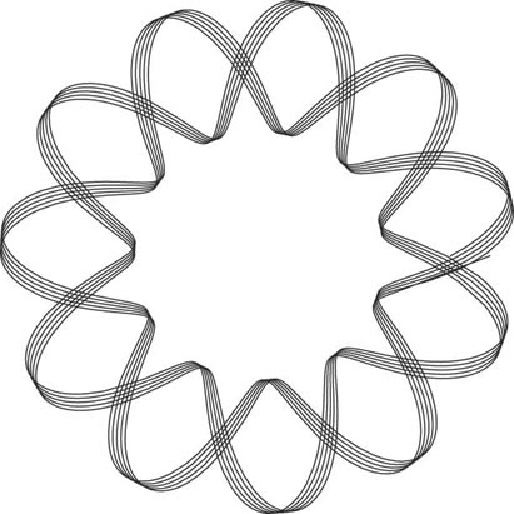}
\caption{\small\
Trace of a slowly precessing, nearly 11:2 resonant vibrotor (compare Fig.~(\ref{fig_spin})), in which the trajectory visits each ``tooth'' of the gear only every other pass. If the resonance had been 11:3, it would be every 3rd pass, etc. This makes the quasiresonance condition easier to break, since the timescale for the transient interaction to now has to be slower by a factor of two compared to an 11:1 resonance.
\label{newflower}}
\end{figure}

The longer the interaction timescale, the more high-order resonances {\it and chaos} can be manifested; see figure (\ref{fig_sig}). Thus, there is a balancing act set up between interaction time and the strength and form of the interaction. By keeping the transient interaction  long compared to the internal time scales associated with the low-order resonances, but short compared 
to the time required to complete the long periods involved in the high-order resonances, we can maximize the low order resonance plateaus.  Similar issues are  faced in the adiabatic switching method \cite{switch}.

\begin{figure}
\centering
\vspace*{0.6cm}
\includegraphics[height=8.9cm]{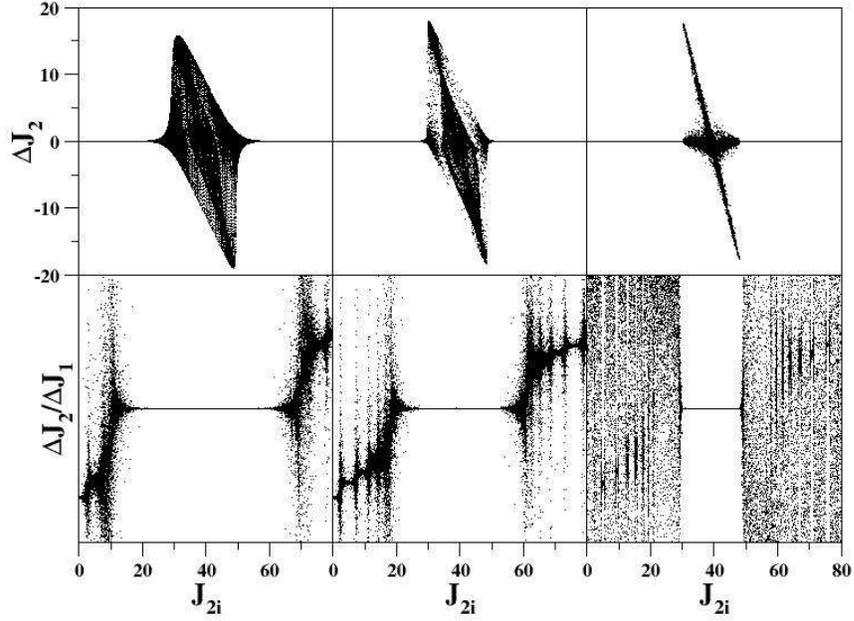}
\caption{\small\
The ratio of action changes and the change in the action $J_{2}$ {\sl vs} the initial action $J_{2i}$
for transient interactions (\ref{poten}) with different time dependence. The panels on the left correspond to 
a Gaussian function (\ref{latt3}) with $\sigma=50$, the panels on the center to $\sigma=100$ and panels 
on the right to $\sigma=500$. In all cases the energy is $E=75$, up to $m=2$ coupling terms have been included 
in the interaction term, which has a strength $v_{0}=0.1$. The different parameters in the unperturbed 
Hamiltonian are the same as in the previous figures. All quantities are expressed in arbitrary units. 
\label{fig_sig}}
\end{figure}
\vskip .1in
\subsection{ Time evolution of the actions in a quasiresonance domain}
\vskip .1in
\begin{figure}
\centering
\includegraphics[height=9.9cm]{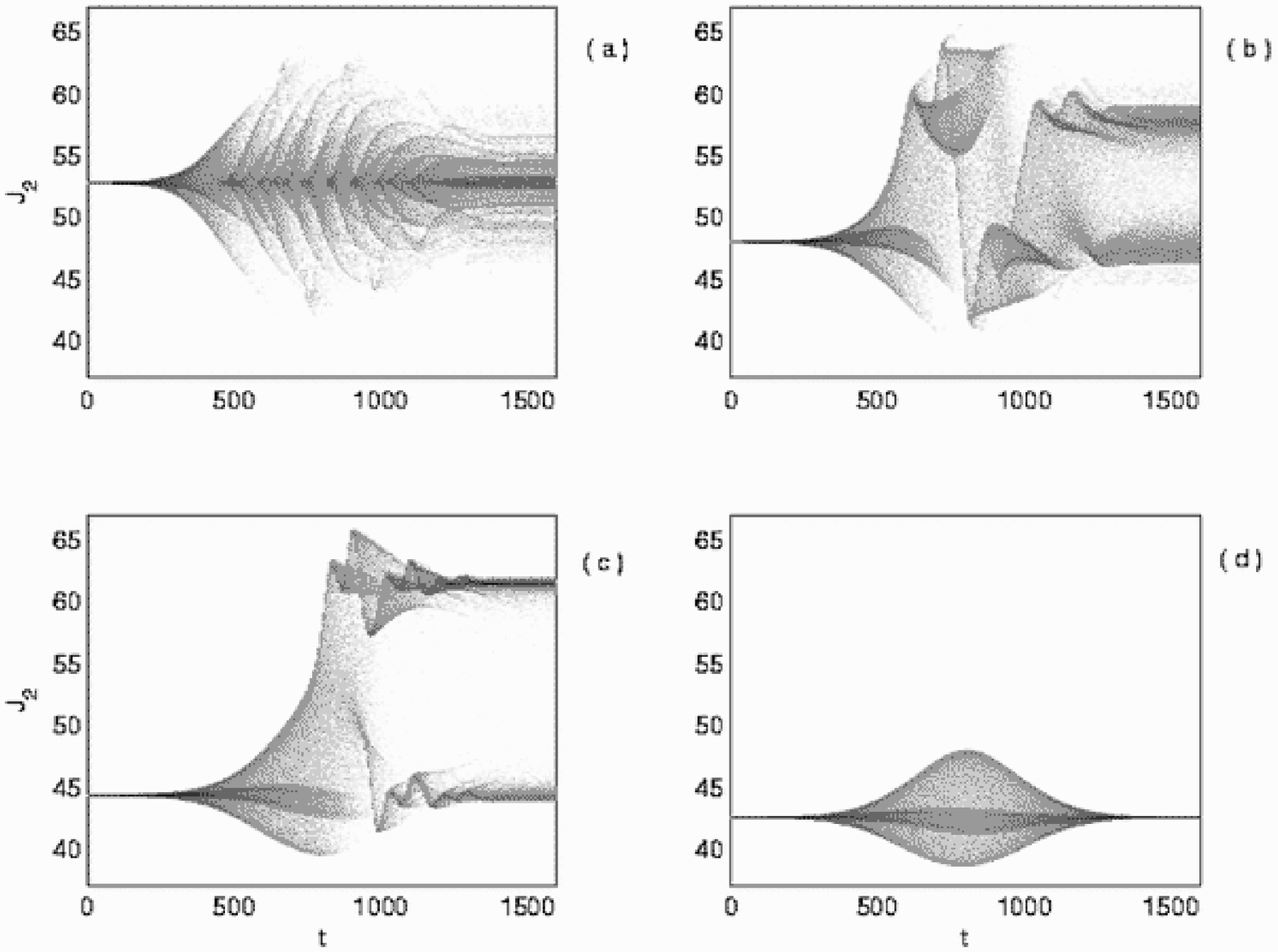}
\caption{\small\ 
The action $J_2$ {\sl vs.} time for an ensemble of $2000$ trajectories with the same initial actions and initial 
angle variables $\Phi_1$ and $\Phi_2$ chosen at random. 
In $(a)$ the initial action is $J_{2i}=52.78$ ($J_{2i}=J_{2,100}^{1:1}$, center of the nonlinear resonance zone),
in $(b)$ is $J_{2i}=48$ ($J_{2,100}^{1:1} < J_{2i} < J_{2,100,1}^{N:M-}$, inside the quasiresonance domain),
in $(c)$ is $J_{2i}=44.3$ ($J_{2i}\simeq J_{2,100,1}^{N:M-}$, one of the edges of the quasiresonance domain) 
and in $(d)$ the initial action is $J_{2i}=42.5$ ($J_{2,100,2}^{N:M-} < J_{2i} < J_{2,100,1}^{N:M-}$,
between the curves of the two separatrices). The amplitude of the transient interaction is maximum at $t=750$. 
In the four panels the same parameters as in figure (\ref{fig_t1}) were chosen. All quantities are expressed
in arbitrary units.
\label{fig_t5}}
\end{figure}

Figure (\ref{fig_t5}) shows the time evolution of an ensemble of trajectories with initial angle variables chosen
at random and four different values of the initial action $J_{2}$. The parameters used are the same as in figure  \ref{fig_t1}, which shows the phase space and action ratio.  In panel $(a)$, where the initial action coincides
with the resonance action $J_{2,E}^{N:M}$, we observe the spread of the action values along a series of branches corresponding 
to the oscillations of the trajectories around the elliptic fixed points when they remain trapped in the resonance zone.
Once the transient coupling interaction starts decreasing and the resonance zone shrinks, the actions settle on different values around the
resonance action and a diffusion in action of the original ensemble is finally observed.

When the initial actions are between the center and the edges of the quasiresonance domain,  
 $J_{2,E}^{N:M} \leq J_{2i} \leq J_{2,E,1}^{N:M\pm}$, see panels $(b)$ and $(c)$, there is a relatively smooth diffusion of the 
 action until the strength of the transient interaction reachs its maximum amplitude. 
Afterwards the ensembles evolve into two horizontal distributions symmetrically arranged with respect to the center of the resonance zone.
The  spread in final action decreases as the initial actions approach to the edges of the quasiresonance domain, $J_{2i} \simeq J_{2,E,1}^{N:M\pm}$.
In this region the ensemble evolves into a bimodal distribution strongly peaked at the final action values  $J_{2f}\simeq J_{2i}$ and
$J_{2f}\simeq -J_{2i}+2J_{2,E}^{N:M}$, which correspond to the two linear branches that give the diagonals of the parallelogram
depicted in figure  (\ref{fig_Jchange}). 

As panel $(d)$ shows, the dispersion of the action is less important for initial values that range from the 
boundary of the innermost separatrix to the outermost separatrix, $J_{2,E,1}^{N:M\pm} \leq J_{2i} \leq J_{2,E,m}^{N:M\pm}$.
Here the distribution of the action values defines a smooth structure that grows when the strength of the transient interaction is 
increasing and then shrinks as soon as such interaction starts decreasing. Thus, when the transient interaction disappears
the action of nearly all the trajectories recover their initial actions. 
Trajectories with initial action even further away, outside the limit of the outermost separatrix curve, $J_{2i} < J_{2,E,m}^{N:M-}$ and 
$J_{2i} > J_{2,E,m}^{N:M+}$, stay outside the phase space region swept by the resonance zone and their  actions remain approximately constant throughout the process. 

\begin{figure}
\centering
\hspace*{-0.5cm}
\includegraphics[height=6.6cm]{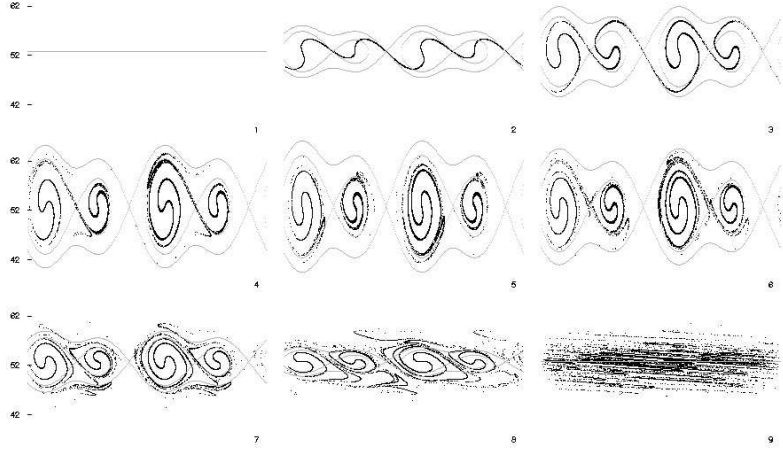}
\caption{\small\ 
The $J_2 - \Phi_2$ phase plane (black dots) for a ensemble of $30000$ trajectories
at different time values. All trajectories have initial angle variables chosen at random and the same initial actions 
as the panel $(a)$ in figure (\ref{fig_t5}). The angle variable ranges from $0$ to $4\pi$. The gray dots represent
the analytical curves of the separatrices (\ref{aver14}) at the same time values. 
The transient interaction is zero in panel $1$, it is increasing from panels $2$ to $4$, it reaches its maximum amplitude
in panel $5$ and it is decreasing from panels $6$ to $9$ where has practically disappeared.
All the parameters of the model are the same as in figure (\ref{fig_t5}).
\label{fig_t6}}
\end{figure}
\begin{figure}
\centering
\hspace*{-0.5cm}
\includegraphics[height=6.6cm]{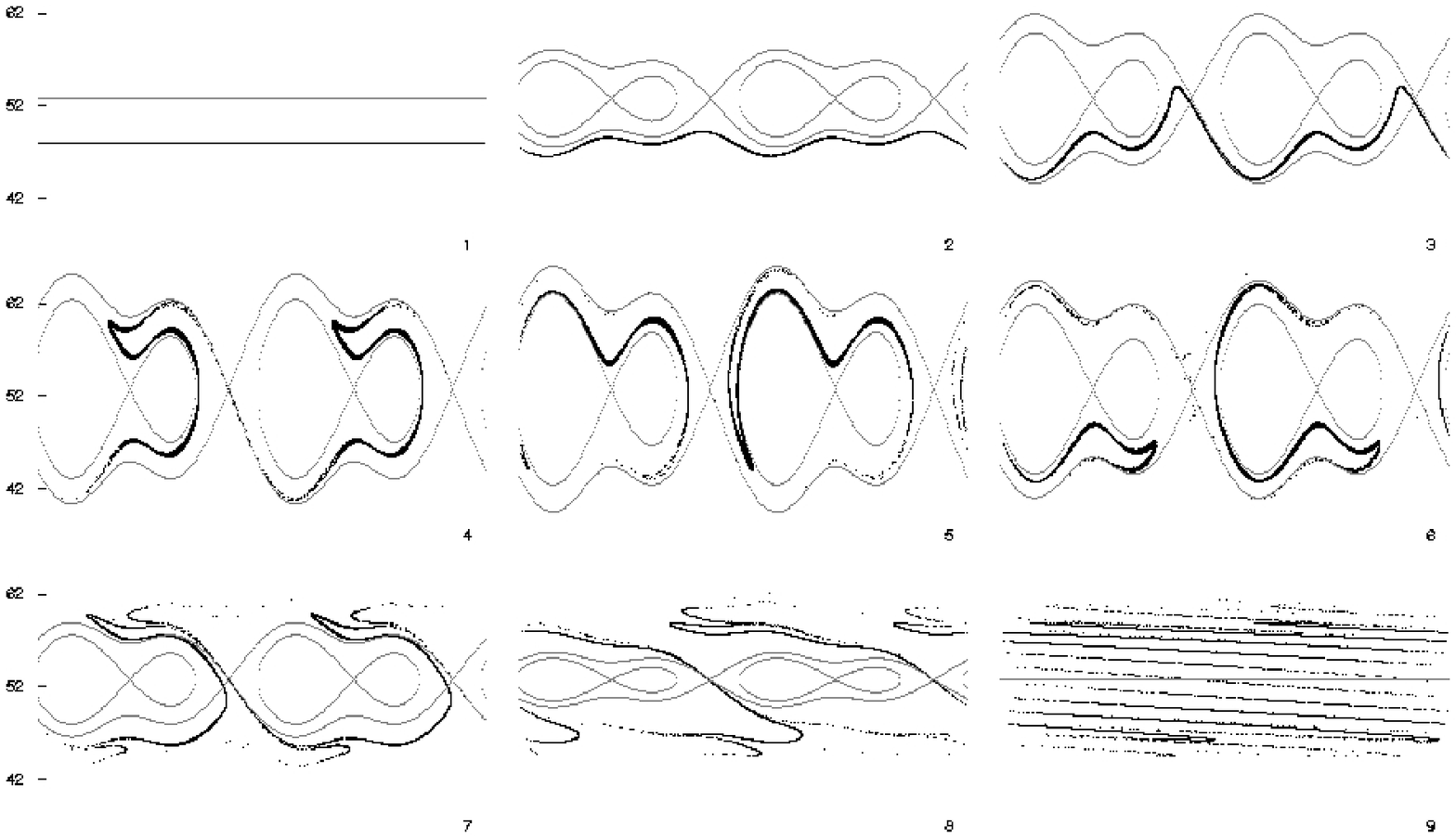}
\caption{\small\ 
The same as figure (\ref{fig_t6}) but for an ensemble with the same initial actions as the panel $(b)$ in 
figure (\ref{fig_t5}).
\label{fig_t7}}
\end{figure}
\begin{figure}
\centering
\hspace*{-0.5cm}
\includegraphics[height=6.6cm]{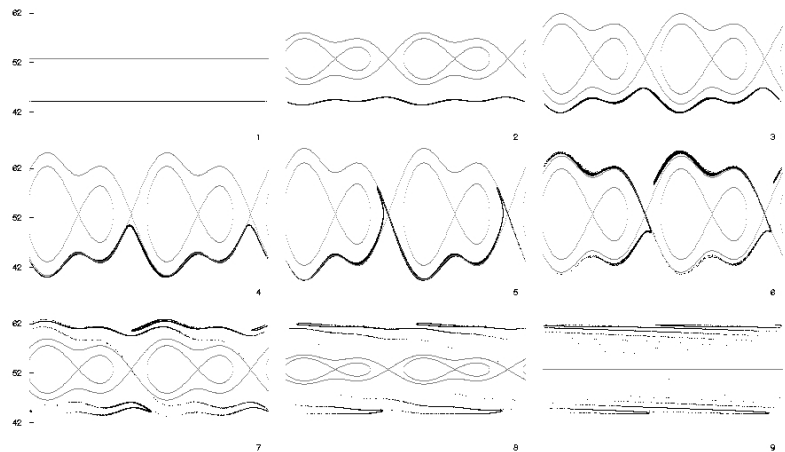}
\caption{\small\ 
The same as figure (\ref{fig_t6}) but for an ensemble with the same initial actions as the panel $(c)$ in 
figure (\ref{fig_t5}).
\label{fig_t8}}
\end{figure}

Figures (\ref{fig_t6}-\ref{fig_t8}) give much insight into the behavior of the actions observed in figure (\ref{fig_t5}) 
when the initial action is within a quasiresonance domain, $J_{2,E,1}^{N:M -} \leq J_{2i} \leq J_{2,E,1}^{N:M +}$.
{\it They show that crossing the separatrices
associated with the nonlinear resonance zone is the underlying mechanism that makes possible the significant changes in the actions
in the quasiresonance effect. }
The breakdown of adiabatic invariance due to the passages through the separatrices of nonlinear resonance zones  is 
a quite general phenomenon which has been observed to play an important role in the dynamics of systems that are 
perturbed by a slowly time-varying interaction; such as the dynamics of charged particles in electromagnetic 
fields in plasma physics, the motion of asteroids in celestial mechanics or the propagation of short radio waves in the ionosphere, 
see \cite{sep_cros} and references therein.
\vskip .1in
\subsection{ Phase space evolution}
\vskip .1in
Phase space pictures add insight to the mechanism of action changes and quasiresonance. Figure (\ref{fig_t6}) displays the evolution in phase space of one ensemble of trajectories with initial actions equal to 
the resonance action, $J_{2i}=J_{2,E}^{N:M}$. Hence, the phase space points of the ensemble are already inside 
the oscillatory region of the resonance zone when it develops from the resonance action $J_{2,E}^{N:M}$.
As the resonance zone grows in time the phase points cannot entirely  escape from this region, but they can cross the 
innermost separatrix curve, getting  trapped in the internal loops and encircling the elliptic fixed points, 
or they can remain confined in the region between the two separatrix curves. 
Figures (\ref{fig_t7}) and (\ref{fig_t8}) include ensembles of trajectories with initial actions between the center and the edges of 
the quasiresonance domain, $J_{2,E,1}^{N:M -} < J_{2i} < J_{2,E,1}^{N:M +}$.
In both sequences the ensembles are initially unperturbed and therefore the action remains nearly constant. But as the resonance zone 
grows, it gradually deforms the invariant tori, $J_2=J_{2i}=const$, that keep the trajectories confined in phase space. 
Once the initial ensemble reaches the outermost separatrix associated with the nonlinear resonance, and
provided that  the resonance zone is still growing in time, some phase space points pass through the separatrix and go into the oscillatory region 
of the nonlinear resonance zone, {\sl i.e.} their dynamics changes from a precession to oscillation. As in the previous sequence, once the phase points are confined in the oscillatory region they can either remain in the transition regions 
between two separatrix curves or go inside the internal loops by passing through the innermost separatrix. 

It is only after the resonance zone starts shrinking that the trajectories  trapped in the oscillatory region can recover the rotational 
dynamics by re-crossing the outermost separatrix. Afterwards there is no way 
 they can pass through the resonance zone again. Since the trajectories can randomly emerge to any point inside the resonance
zone, the final effect of crossing the separatrices is the diffusion in action of the original ensemble around the resonance action value. 
As the transient interaction goes away, the destroyed unperturbed tori, $J_i=const\,(i=1,2)$, are replaced by the {\sl modified} invariant tori, 
$I_2=const$, that characterize the quasiresonance effect. These new invariant tori confine the trajectories across the phase space region
previously swept by the separatrices of the nonlinear resonance zone. Since such region ranges from $J_{2,E,1}^{N:M -}$ to $J_{2,E,1}^{N:M +}$, 
the length of the quasiresonance domain, important variations of the actions may occur after the transient interaction goes away.

As we mentioned before, the nonadiabatic effects that arise in the proximity of the edges of the quasiresonance domains are associated with the 
progressive approach of the trajectories in this region to the boundary of the nonlinear resonance zone. As figure (\ref{fig_t8}) shows, these are
trajectories that intersect the separatrices very close to the hyperbolic fixed points, where the motion of the system becomes extremely slow, and 
no adiabatic invariant remains. 
Trajectories with initial actions inside a well defined quasiresonance domain, see figure (\ref{fig_t7}), intersect the separatrices far enough from 
the hyperbolic points and do not generate significant nonadiabaticity. 
These trajectories can go deeper into the oscillatory region of the resonance zone and fully perform the quasiperiodic motion around the elliptic fixed 
points that gives the quasiresonance effect.
But, the more slowly the transient interaction proceeds, the more trajectories get closer to the hyperbolic fixed points when the crossings of the 
separatrices occur and, consequently, the more extended across the quasiresonance domain the nonadiabaticity effects manifest.

The increasing attraction of the trajectories with initial action inside the quasiresonance domains towards the unstable fixed points is clearly observed 
in figure (\ref{fig_pend}), which shows the evolution of two trajectories of the pendulum Hamiltonian (\ref{aver10}) with the same initial action 
but different slowness of the transient interaction. 
\begin{figure}
\centering
\hspace*{-0.5cm}
\includegraphics[height=6.3cm]{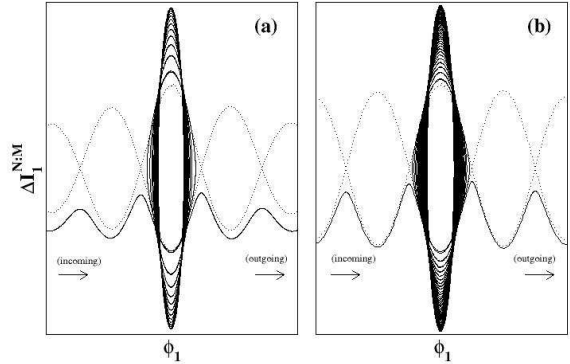}
\caption{\small\ 
Time evolution of two trajectories of the  pendulum Hamiltonian (\ref{aver10}) with different Gaussian parameters in the time-dependent function
$g(t)$; $\sigma=1000$ in panel $(a)$ and $\sigma=4000$ in panel $(b)$. Both trajectories had the same initial {\sl action}, inside a quasiresonance domain. 
The dotted line gives the evolution of the separatrix. Note the closer approach of the trajectories to the separatrix  in the case of the longer interaction time.
\label{fig_pend}}
\end{figure}
Initially the pendulum is rotating and the angle variable is a monotonically increasing function, but as the strength of the transient interaction increases
 the rotation slows down. The more slowly the transient interaction evolves the more the trajectory sticks to the hyperbolic fixed point. When 
 the energy of the separatrix equals the energy of the pendulum, the trajectory crosses the separatrix and the transition to the oscillatory regime occurs. 
 Afterwards the pendulum starts orbiting around the elliptic fixed points describing large loops that cover an extensive region in phase space. Later, when the 
 strength of the transient interaction is decreasing, a new intersection of the separatrix occurs and the pendulum recovers its rotational motion. The more slowly
  the transient interaction proceeds the closer to the hyperbolic fixed point the trajectory emerges from the oscillatory region.  
Thus, in the limit of extremely slowly varying transient interaction, when the closest approach to the hyperbolic fixed points occurs, the outgoing trajectories
from the oscillatory region of the resonance zone asymptote to the two unstable manifolds that emanate from such points, see figure (\ref{fig_pend2}).
It is the evolution along such manifolds which  gives the bimodal distribution in the final action values previously discussed.
 
\begin{figure}
\centering
\hspace*{-0.5cm}
\includegraphics[height=6.3cm]{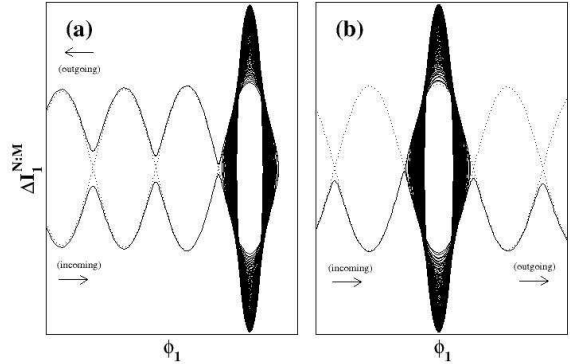}
\caption{\small\ 
Evolution of two trajectories of the pendulum Hamiltonian (\ref{aver10}) in a extremely slowly varying transient interaction, $\sigma=10000$. 
In $(a)$ the outgoing trajectory from the oscillatory region follows the unstable manifold that gives the branch $\Delta J=-2J_{i}+2J_{c}$
in the bimodal distribution depicted in figure (\ref{fig_Jchange}), and in $(b)$ the trajectory evolves along unstable manifold that gives the branch 
$\Delta J=0$. Both trajectories have the same initial action, but different angle variable. The dotted line is the separatrix.
\label{fig_pend2}}
\end{figure}

\section{Conclusions}
\label{concl}
We have shown that quasiresonance is a common effect, not restricted to vibro-rotationally inelastic
collisions between an atom and a diatomic molecule, which may arise in any classical integrable system
that is perturbed by a transient interaction with additional degrees of freedom which couples {\sl quasi}
(not necessarly exactly) resonant internal degrees of freedom. We have also shown that quasiresonance effects
arise from either an autonomous or an explicitly time dependent transient interaction.

The analysis of the variations of the actions and the energy of the system in a quasiresonance domain  
reveals that the ratio of action changes is paramount, adhering extremely closely to a rational ratio involving small integers, with the energy playing a secondary role. 

In the limit of weak and slowly varying transient interaction, the classical adiabatic invariance theory and the method of 
averaging have been used to establish the link between the presence of a quasiresonance domain and the existence 
of a topologically modified adiabatic invariant that characterizes the quasiperiodic motion of the system in the inner oscillatory 
region of a nonlinear resonance zone in phase space. Within a resonance zone, the rational value in the ratio of action changes that 
gives a quasiresonance follows immediately from the linear integer combination that defines the modified adiabatic invariant, i.e. the 
action of the fast motion, in terms of the action variables of the unperturbed system.  Outside the resonance zones there is practically
 no response from the system to the perturbative interaction and the adiabatic approximation is good for both unperturbed actions, but the ratio of the tiny changes is not remarkable.

The generalized {\sl pendulum} Hamiltonian obtained from the averaging over the fast motion in the proximity of a resonance zone 
has provided a good model to clarify the dynamics of the system when a quasiresonance occurs. The analysis of the phase space of such 
Hamiltonian has shown that the maximum size of the phase space region swept by the innermost separatrix of the nonlinear resonance in the
 course of time gives a good estimate for the length of the corresponding quasiresonance domain. 
This analysis has also shown that the change in the dynamics of the system due to the crossing of the separatrices as they move in  
phase space is the mechanism involved in the significant variations of the actions observed in a quasiresonance. 

A natural extension of the work presented in this paper is the generalization of the concept of quasiresonance to 
systems with more than two or three dimensions. To approach such systems, we can return to the physical picture of the averaged Hamiltonian, and the nature of the interaction, where the averaging is over all  fast variables which remain after identifying the slow ones (we do not assert that this is always possible).  Whenever the transient interaction timescale (autonomous or not) is slow compared to the  time between successive visits to the interaction region, the quasiresonance concept may survive the extension to many dimensions. There may remain  more than one slow variable, in which case the nonlinear evolution will take place in the slow variable space, perhaps preserving none of their individual actions. Nonetheless all the fast actions would be conserved. 

Another extension of this work  is its connection with possible models of Arnol'd diffusion\cite{lich_lieb}. The picture would be of random but quasiresonant travel along resonance surfaces in action space, interrupted by non-quasiresonant but extremely small action changes at the edges of quasiresonant plateaus.

\section{Acknowledgments}
This paper is submitted in honor of and in the spirit of the many deep contributions of  Prof. Mark Child to classical physics associated with molecular and quantum systems.

A.R. would like to thank Robert Parrot, Diego V. Bevilaqua and Daniel Alonso for many fruitful discussions.
A.R. gratefully acknowledges the hospitality of the Physics Department of Harvard University, where the 
present work was carried out.
A.R. was supported by a Fulbright/MECD grant from Secretar\'\i a de Estado de Educaci\'on 
y Universidades and, partially, from Fondo Social Europeo. This work was also supported by a grant from the National Science Foundation, NSF CHE-0073544.

\end{document}